\documentclass[10pt,journal,compsoc]{IEEEtran}
\IEEEoverridecommandlockouts

\usepackage{adjustbox}
\usepackage{subfigure}
\usepackage{mathtools}
\usepackage{balance} 
\usepackage{url}
\usepackage{slashbox}
\usepackage{graphicx}
\usepackage{caption}
\usepackage{multirow}
\usepackage{array}
\usepackage{makecell}

\newcolumntype{L}[1]{>{\raggedright\let\newline\\\arraybackslash\hspace{0pt}}m{#1}}
\newcolumntype{C}[1]{>{\centering\let\newline\\\arraybackslash\hspace{0pt}}m{#1}}
\newcolumntype{R}[1]{>{\raggedleft\let\newline\\\arraybackslash\hspace{0pt}}m{#1}}

\ifCLASSINFOpdf
\else
\fi

\begin{document}

\title{\Large \bf iTeleScope: Intelligent Video Telemetry and Classification in Real-Time \\using Software Defined Networking}

\author{Hassan~Habibi~Gharakheili,
			Minzhao~Lyu,
			Yu~Wang,
			Himal~Kumar,		
			and~Vijay~Sivaraman
			
	\IEEEcompsocitemizethanks{\IEEEcompsocthanksitem H. Habibi Gharakheili, M. Lyu, H. Kumar, and V. Sivaraman are with the School of Electrical Engineering and Telecommunications, University of New South Wales, Sydney, NSW 2052, Australia (e-mails: h.habibi@unsw.edu.au, minzhao.lyu@unsw.edu.au, yu.wang1@unswalumni.com, 
		himal.kumar@unsw.edu.au, vijay@unsw.edu.au).
	\protect\\
	\IEEEcompsocthanksitem This article is an extended and improved version of our paper presented
	at the EWSDN 2016 conference \cite{ewsdn16}.	
	}

}


\IEEEtitleabstractindextext{%
	\begin{abstract}
		Video continues to dominate network traffic, yet operators today have poor visibility into the number, duration, and resolutions of the video streams traversing their domain. Current approaches are inaccurate, expensive, or unscalable, as they rely on statistical sampling, middle-box hardware, or packet inspection software. We present {\em iTelescope}, the first intelligent, inexpensive, and scalable SDN-based solution for identifying and classifying video flows in real-time. Our solution is novel in combining dynamic flow rules with telemetry and machine learning, and is built on commodity OpenFlow switches and open-source software. We develop a fully functional system, train it in the lab using multiple machine learning algorithms, and validate its performance to show over 95\% accuracy in identifying and classifying video streams from many providers including Youtube and Netflix. Lastly, we conduct tests to demonstrate its scalability to tens of thousands of concurrent streams, and deploy it live on a campus network serving several hundred real users. Our system gives unprecedented fine-grained real-time visibility of video streaming performance to operators of enterprise and carrier networks at very low cost.
	\end{abstract}
	
	\begin{IEEEkeywords}
		SDN, Telemetry, Calssification, Video Traffic.
	\end{IEEEkeywords}
}

\maketitle
\IEEEdisplaynontitleabstractindextext
\IEEEpeerreviewmaketitle

\IEEEraisesectionheading{\section{Introduction}\label{sec:intro}}

\IEEEPARstart{V}{ideo} constitutes a majority of Internet traffic today, and is slated to increase even further in the near future, as higher resolutions (1440p and 4K) become more prevalent, and augmented/virtual reality (AR/VR) starts to take off \cite{ciscoVNI16}. In order to manage this video traffic (for quality and cost reasons), enterprises and carriers need better visibility into the video streams in their network. Operators can today infer macro-scopic attributes (such as aggregate volume of video traffic on their peering link with a video content provider like Netflix), but they have near-zero visibility into the micro-scopic aspects, such as how many video streams are concurrently active at a time, what their durations are, what resolutions they operate at, and how often they adapt their rate. Visibility into these attributes can allow them to better understand both video content characteristics and video viewing patterns, so they can tune their network to meet content-provider expectations as well as enhance user experience.

Several existing methods can be used for visibility into video streams, but they come with disadvantages: SNMP \cite{SNMP} can be used to retrieve traffic counts from switches, but these counters are at the interface-level, and may represent an aggregate of many video and non-video flows. NetFlow \cite{NetFlow} enables a switch to aggregate IP flow information in a local cache and export this information periodically -- this requires the switch hardware to be capable of decoding, collating and caching entries, and can also entail a penalty in switch CPU utilization in the range of 7-22\% \cite{NetFlowPerf}. sFlow \cite{sFlow} reduces this overhead by statistically sampling traffic; however, lower sampling rates inevitably lead to reduced accuracy in traffic characterization. Specialized traffic monitoring solutions can provide both accuracy and performance -- for example deep packet inspection ``middle-boxes'' (e.g. Sandvine) can inspect data packets at high rates; however, such solutions cost hundreds of thousands of dollars that is prohibitive for many network operators.

The general problem of classifying network traffic has been studied by many prior research works \cite{trafficClass12,taxonomyTC13,dpiSurvey14,FlowQoS14,MLtrafficClass15, Nmeta15,vTC16}, using various methods ranging from inspecting a few bytes in the payload, to processing headers or characterizing the signatures of packets streams. Our paper focuses more narrowly on streaming video flows, and the general methods developed earlier do not directly apply, as they are either reliant on packet payloads being visible (video traffic is increasingly encrypted) or require long trains of packets to be analyzed in software (limiting scalability). Further, they do not determine aspects specific to video streams, such as rates and resolutions. We believe that the Software Defined Networking (SDN) paradigm is ideally suited to the task of identifying and classifying video traffic, since Openflow by its nature provides flow-level isolation and visibility in a low-cost and scalable manner.

While SDN-based flow-level monitoring for video streams may seem conceptually simple, there are several challenges to be overcome: {\em correctness} of the solution requires dealing with an arbitrary set of content providers and dynamic video end-points; carrier-grade {\em performance} requires the controller to be protected against packet overload and the solution to be resilient to controller failures; and high {\em scalability} requires minimizing the software inspection of packets as well as flow-modifications on the switch hardware. In this paper we develop, deploy, and evaluate our SDN-based solution called \textit{iTeleScope} that meets these challenges to provide fine-grained visibility into streaming video flows. Our {\bf first} contribution is to develop a system architecture, comprising selective packet inspection, dynamic flow-table management, and flow traffic profile analytics, that is intelligently able to identify and classify long video streams. We show how our design meets our goals of low cost (by using commodity Openflow switches), scalability (by filtering packets to minimize software processing), and high accuracy (via machine learning methods that use key attributes related to flow traffic profiles). Our {\bf second} contribution develops a fully-functional prototype based on the above architecture using commodity hardware and software: a NoviFlow Openflow switch, the Ryu SDN controller, the Bro packet inspection engine, and the Weka machine learning suite. We show how we train and tune our classifier in the lab, and validate its performance to obtain over 95\% accuracy in identifying video streams and deducing their resolution, typically within 60-90 seconds, even when such traffic comes from shared server pools that serve many types of content (such as done by Google). {\bf Finally}, we demonstrate the scalability of our system to tens of thousands of concurrent streams generated from a hardware tester, and do a field-deployment in a University dorm network serving hundreds of students, yielding new insights into video viewing patterns and quality measures for the University residence network.

The rest of this paper is organized as follows: \S\ref{prior} describes prior work on network monitoring solutions, and \S\ref{arch} describes our solution approach that captures and evaluates flow-level information. In \S\ref{proto} we describe our prototype implementation used to validate our solution, while in \S\ref{eval} we evaluate the scalability and efficacy of our system. The paper is concluded in \S\ref{con}.

\section{Related Work}\label{prior}

\textbf{Traffic classification}: This has been a broad-ranging area of interest to the research community for well over a decade, aiming to distinguish mice from elephants, peer-to-peer from downloads, and over-the-top voice/video from streaming applications. Several surveys of this area have been conducted \cite{trafficClass12,taxonomyTC13, dpiSurvey14} and reveal existing classification techniques to have different trade-offs in terms of their \textit{accuracy, computing cost, and scalability}. Among widely used approaches: (a) TCP/UDP port-based classifiers have become less reliable since modern sophisticated applications use non-standard or random port numbers; (b) payload inspectors come at high cost of processing and are increasingly being defeated due to encrypted content; (c) statistical and behavioural classifiers are attractive as they are fairly light-weight, employing flow-level information and machine learning algorithms to identify various traffic types. It has been shown \cite{NetflowTutorial14}  that flow-level analysis (i.e. based on NetFlow and IPFIX) is more scalable than packet-level analysis, and combination of packet inspection and flow monitoring achieves more accurate results.
Unlike much of the prior work that tries to classify traffic by application type, our focus in this paper is specifically on streaming video for reasons stated earlier. As we will see soon, our solution uses a combination of packet filtering based on header information, flow-level telemetry, and behavioral pattern recognition to detect and classify streaming video flows in real-time.

\begin{figure*}[t!]
	\begin{center}
		\mbox{
			\subfigure[iTeleScope system]{
				{\includegraphics[width=0.46\textwidth]{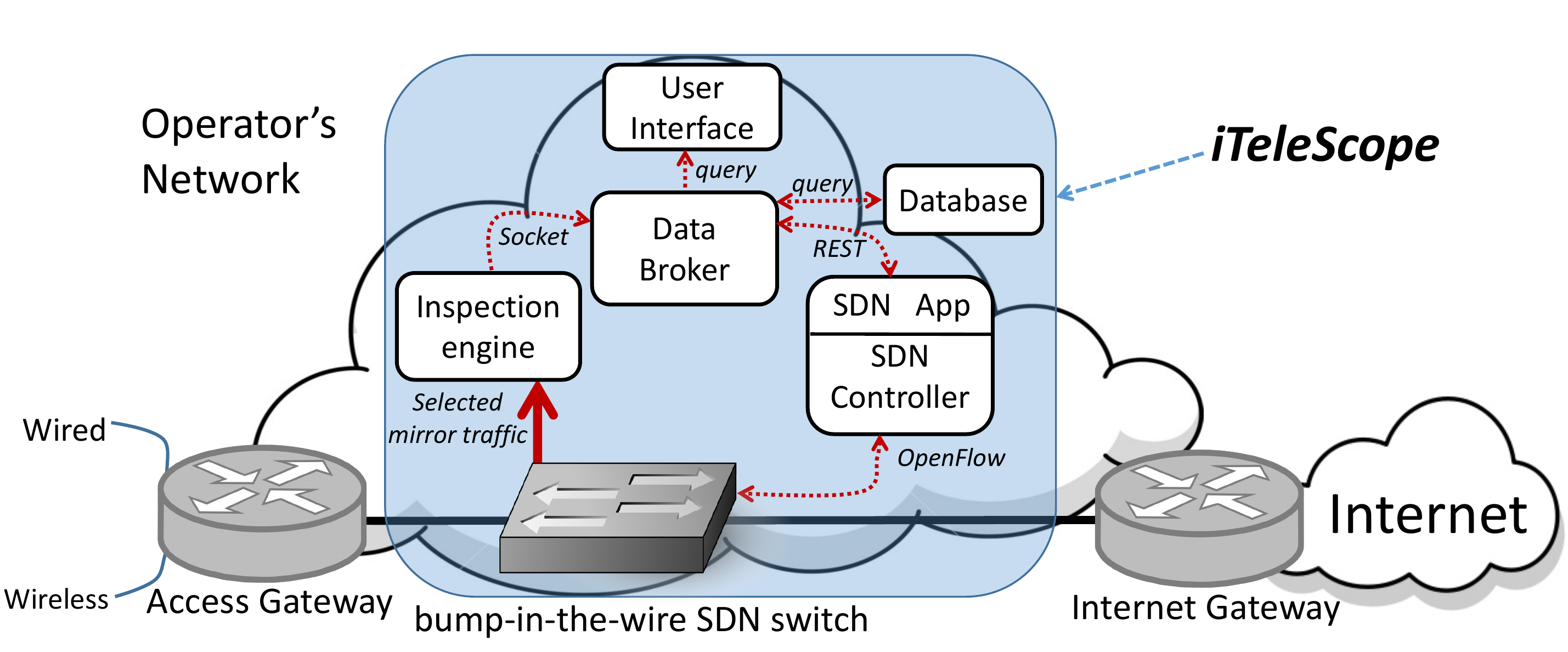}}\quad
				\label{fig:arch}
			}
		}
	\hspace{-2mm}
		\mbox{
			\subfigure[Data Broker]{
				{\includegraphics[width=0.25\textwidth]{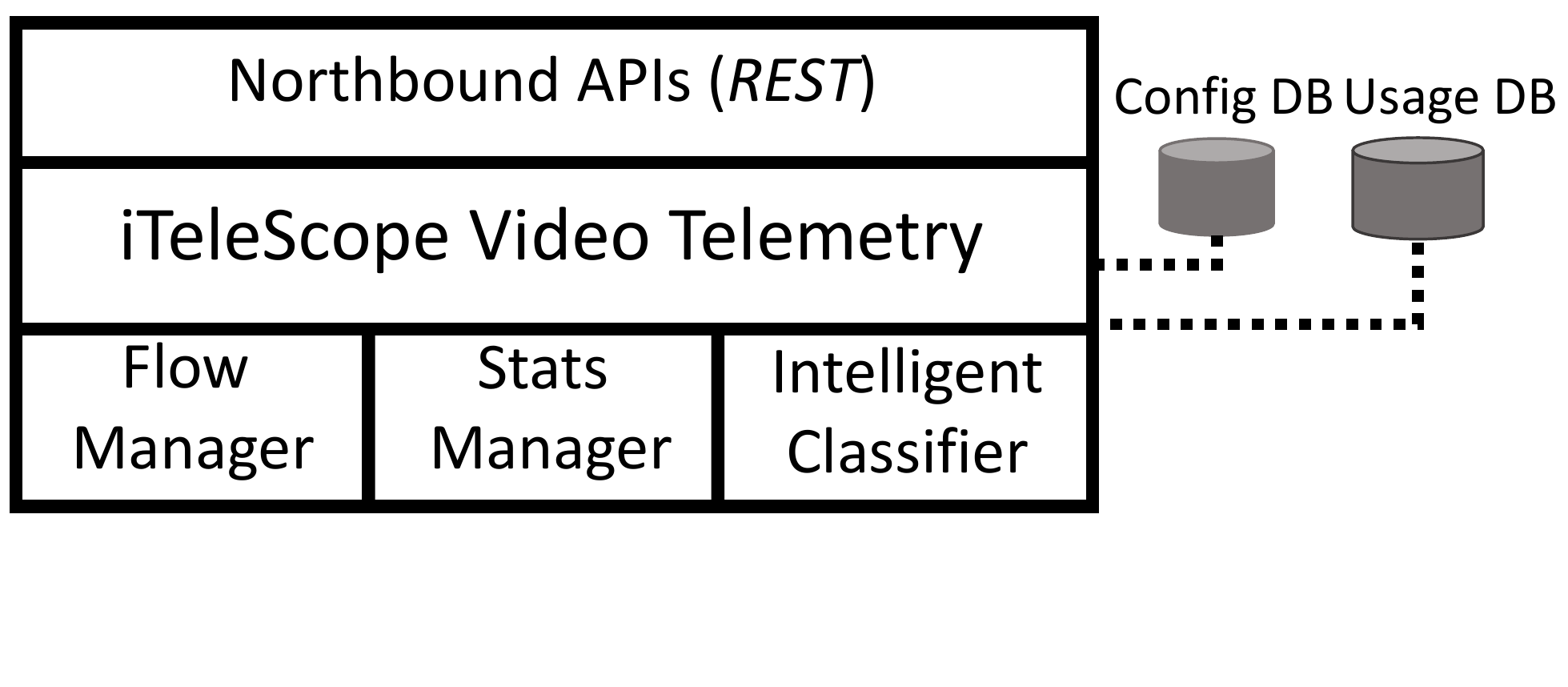}}\quad
				\label{fig:broker}
			}
			\hspace{-2mm}
			\subfigure[SDN application]{
				{\includegraphics[width=0.18\textwidth]{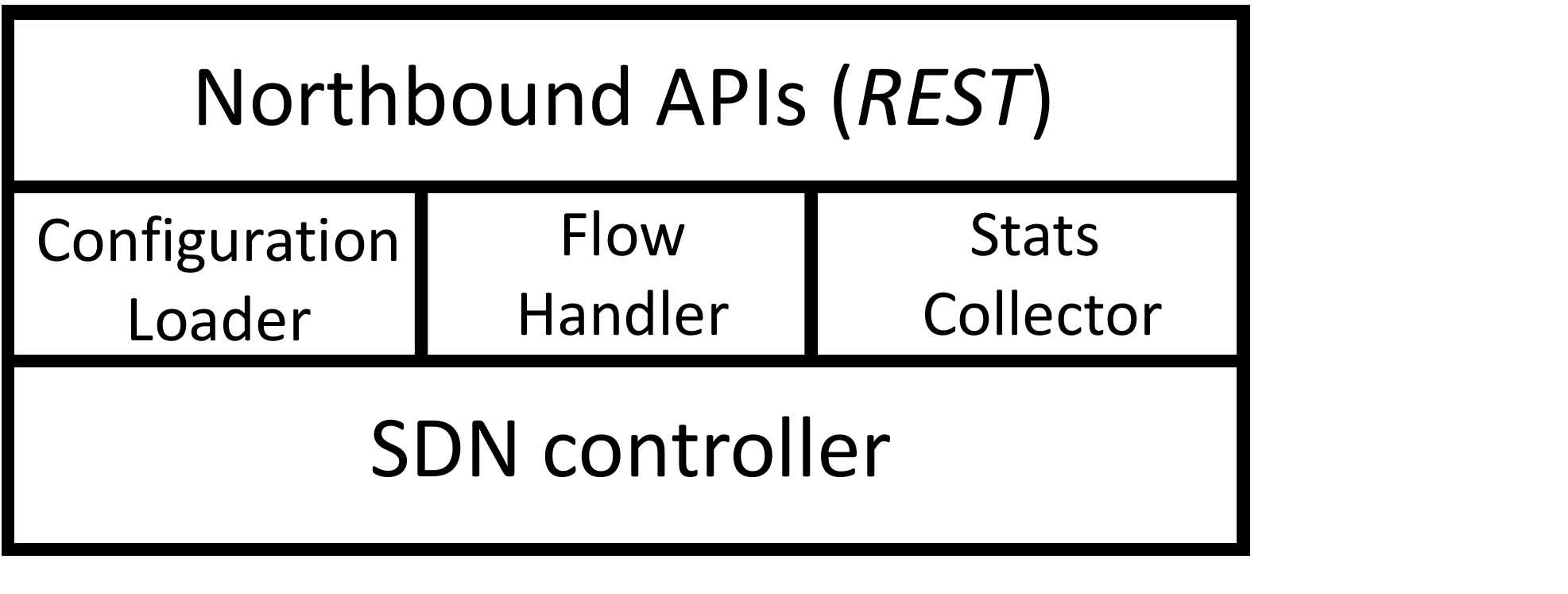}}\quad
				\label{fig:app}
			}
		}
		\vspace{-4mm}
		\caption{iTeleScope (a) system, (b) data broker, and  (c) SDN application, architecture. }
		\vspace{-3mm}
		\label{fig:iTelescope}
	\end{center}
\vspace{-3mm}
\end{figure*}


\textbf{SDN-based monitoring}: Several proposals for flow-based measurement and monitoring for SDN have been developed in the literature \cite{DevFlow11, OpenSketch13, PayLess14, FlowSense13, OpenNetMon14}. OpenSketch \cite{OpenSketch13} proposes a clean slate redesign of the data-plane to support monitoring; unfortunately it requires an upgrade to the data-plane, which can be a barrier to uptake. DevoFlow \cite{DevFlow11} highlights the performance limits of OpenFlow when scaling to extract counters for all flow entries. We partially avoid these problems by inserting reactive entries for only a small fraction of flows, namely flows that transfer a large volume of traffic. Examples of standard OpenFlow based approaches for traffic monitoring include PayLess \cite{PayLess14}, FlowSense \cite{FlowSense13}, Sample\&Pick \cite{Sigcomm15} and OpenNetMon \cite{OpenNetMon14}, which insert rules and collect per-flow counters in response to standard \verb|PacketIn| and \verb|FlowRemoved| messages respectively. These approaches try to balance the accuracy of flow-level statistics against the cost of control-plane overhead, by adjusting the time-out attribute of rules for example based on their counters (a large byte-count shortens the time-out). However, we believe that these reactive interactions between the controller and the switch (i.e. \verb|PacketIn| and \verb|FlowRemoved|) can impose a heavy processing load on the controller, and suffer from disruptions in data-plane operation if there is a control-plane failure. As we will explain later, our architecture does not use any \verb|PacketIn| messages, minimizing the use of controller resources, while also being robust to failures.

The work in \cite{IntMon15} proposes an interactive user interface that uses SDN to monitor and visualize the network state, including traffic rates and rules within each switch. The network administrator can adjust the time-outs of rules as well as the frequency of statistics collection. Our solution also provides an intuitive web-based user interface, though it is specifically for visualizing video flows, and it manages flow-table entries automatically without operator involvement. The work in \cite{flowMeasure16} conducts empirical studies to show that flow-level counters in OpenFlow switches may have significant inaccuracies; however, we have confirmed in our experimental work (\S\ref{eval:perf}) that our NoviFlow switches are accurate to within $1.7$\% in terms of their flow byte-counts.

The current work builds upon our earlier proposal for flow-based video telemetry \cite{ewsdn16}. However, our earlier work relied only on average bitrate, with threshold detectors statically configured for selected video content providers. The current work extends it by extracting a richer set of attributes (such as idle-fraction and burstiness at various time-scales) and using machine learning to automatically identify and classify video flows. Further, we expand our system to incorporate DNS inspection to identify a larger set of content providers, and validated its performance in a live network serving several hundred users.


\section{System Design and Architecture}\label{arch}
In this section we describe our {\em iTelescope} solution, including the major architectural decisions (\S\ref{arch:choice}), the functional blocks (\S\ref{arch:scn}), flow-table management (\S\ref{arch:rule}), packet inspection (\S\ref{arch:anal}), telemetry collection (\S\ref{arch:algo}), and the classification algorithm (\S\ref{arch:ML}).

\subsection{Architectural Decisions} \label{arch:choice}
Our solution is designed to be a ``bump-in-the-wire'' on the link at which video classification is desired (an alternative approach is to feed our system a mirror of all organistional traffic, though this precludes active traffic management at a later date). Our system is therefore transparent to the network, and does not modify packets in any way. Further, our SDN switch does not send any data packets to the controller; instead, packets that need to be inspected in software are sent as copies on a separate interface of the switch, to which a software inspection engine is attached. This protects the controller from overload from the data-plane, allowing it to scale to high rates and to service other SDN applications. Moreover, since incoming data packets are sent onwards by the switch immediately, the data-plane benefits in having minimal latency overhead, and is protected from controller failures.

Our second architectural decision is in the judicious combination of packet-level and flow-level monitoring. In essence, we use the Openflow switch as a hardware filter to limit the fraction of traffic (to the first few MB of each flow) that is mirrored for software inspection; heavier (elephant) flows are suppressed from software mirroring by inserting reactive flow-table entries, and are monitored by periodically polling their flow-level counters. This approach is tuned to balance the load between the software (for packet inspection) and the hardware (for flow table size, modification rate, and counter polling). We undertook an experimenting investigation of this trade-off using a 24-hour feed of campus traffic from a large University, and found that a volume threshold of $4$ MB is suitable for declaring a flow as an elephant and creating a reactive flow-table entry for it. With this threshold, we found that less than 1\% of flows were elephants (requiring less than 5000 entries in the hardware flow-table at any time and less than 20 flow-mods per-second), while no more than one-third of packet traffic was sent to the software inspection engine (since 70-75\% of the traffic was carried in elephant flows). As we will demonstrate later, this balance between hardware and software processing reduces cost and increases scalability, while enabling extraction of attributes for machine learning based classification with high accuracy.


\begin{figure}[!t]
	\hspace{-3mm}
	\includegraphics[width=.5\textwidth]{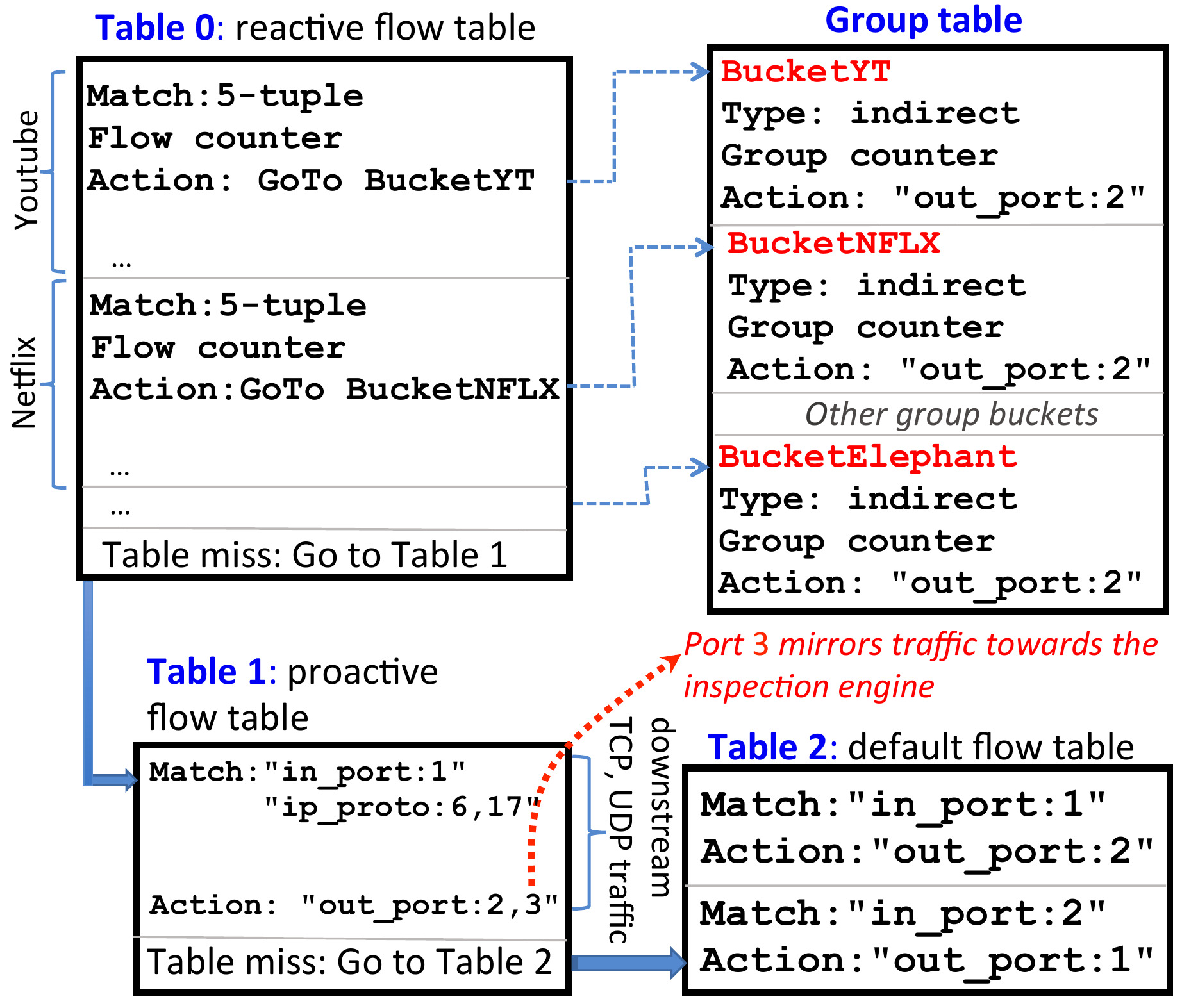}
	\caption{Flow table structure}
	\label{fig:flowtable}
\end{figure}

\subsection{Functional Blocks}\label{arch:scn}
Fig.~\ref{fig:arch} shows the functional blocks in the iTeleScope architecture applied to a typical carrier or enterprise network. End-users are on the left, and can be on an access network using wired (DSL, Ethernet, Fiber) and/or wireless (e.g. 3G/4G, WiFi) technology. The video content providers are on the right, connected to the carrier/enterprise network through an Internet gateway. Our iTelescope solution can be applied on any desired link as a bump-in-the-wire. It comprises an SDN switch whose flow-table rules will be managed dynamically (explained in \S\ref{arch:rule}), a packet inspection engine (described in \S\ref{arch:anal}), and a data broker in conjunction with our App on the SDN controller (internal modules shown in Figures~\ref{fig:broker} and~\ref{fig:app}) that collect telemetry (\S\ref{arch:algo}) and run the classification algorithms (\S\ref{arch:ML}).

The operational flow of events is as follows: assume that video traffic enters (from the content provider) on port-1 and exits (towards the consumer) on port-2; the switch is initially configured to mirror all traffic to the inspection engine on port-3. The inspection engine keeps track of flow volume, and once it exceeds the threshold of 4 MB, notifies the data broker. The broker then instructs our SDN App to insert a reactive flow-entry for the specific stream, which stops the mirroring of packets for this stream. Thereafter, the data broker polls the counters (via SDN App) periodically and develops a traffic profile for the stream, which is fed to the machine learning algorithm for classification. Flow entries for a stream are automatically aged out upon inactivity.
Further, our inspection engine has a specific event handler that captures DNS A-type replies, and extracts the server name and IP address so the content provider for the video stream can be identified. In what follows we describe each of the components in more detail.


\begin{figure*}[t!]
	\begin{center}
		\mbox{
		        \hspace{-4mm}
			\subfigure[Youtube 144p(low resolution)]{
				{\includegraphics[width=0.24\textwidth]{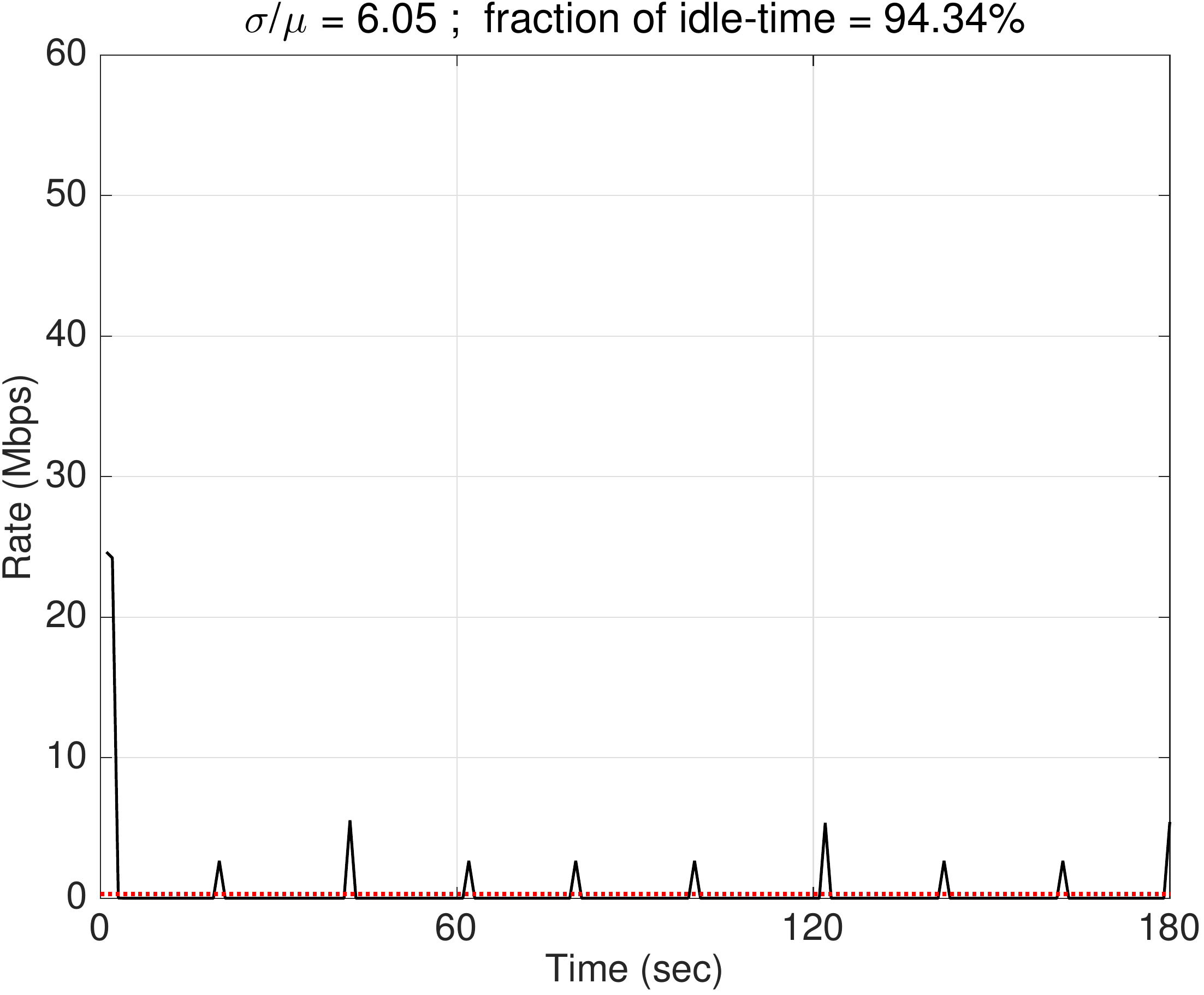}}\quad
				\label{fig:prof144}
			}
			
			\hspace{-4mm}
			\subfigure[Youtube 480p(medium resolution)]{
				{\includegraphics[width=0.24\textwidth]{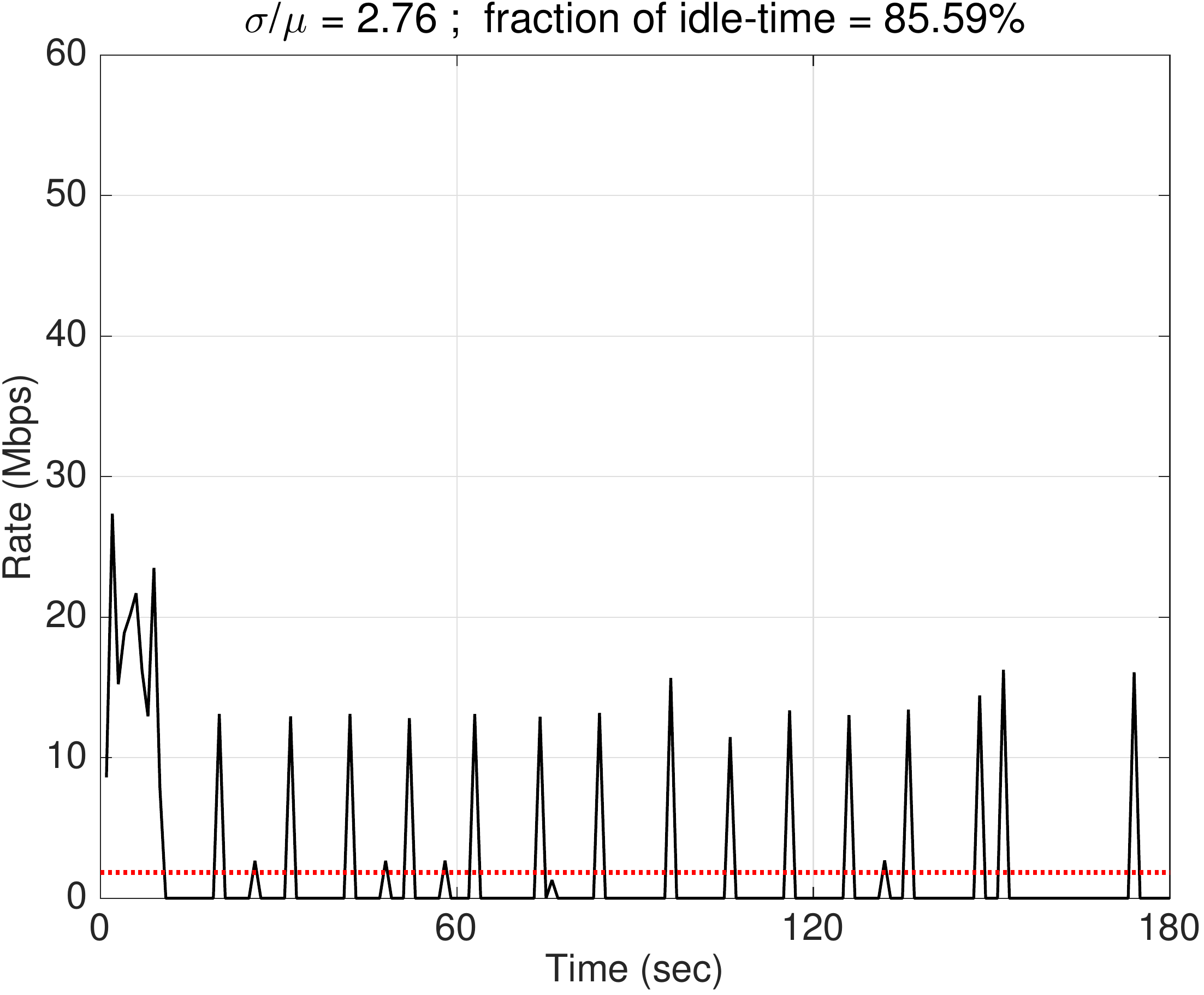}}\quad
				\label{fig:prof480}
			}
			\hspace{-4mm}			
			\subfigure[Youtube 1080p(high resolution)]{
				{\includegraphics[width=0.24\textwidth]{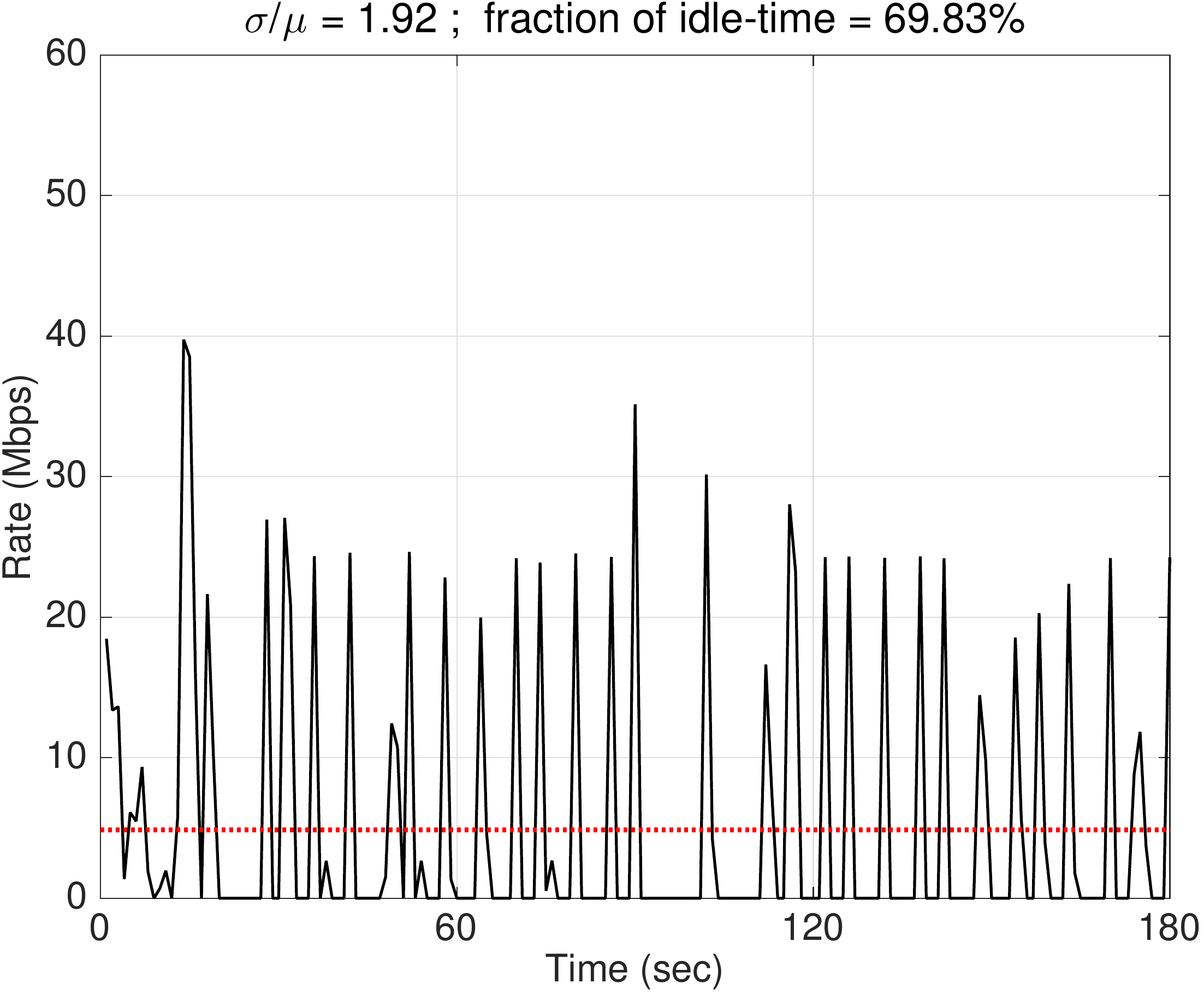}}\quad
				\label{fig:prof720}
			}
			\hspace{-4mm}			
			\subfigure[Youtube 2160p(u-high resolution)]{
				{\includegraphics[width=0.24\textwidth]{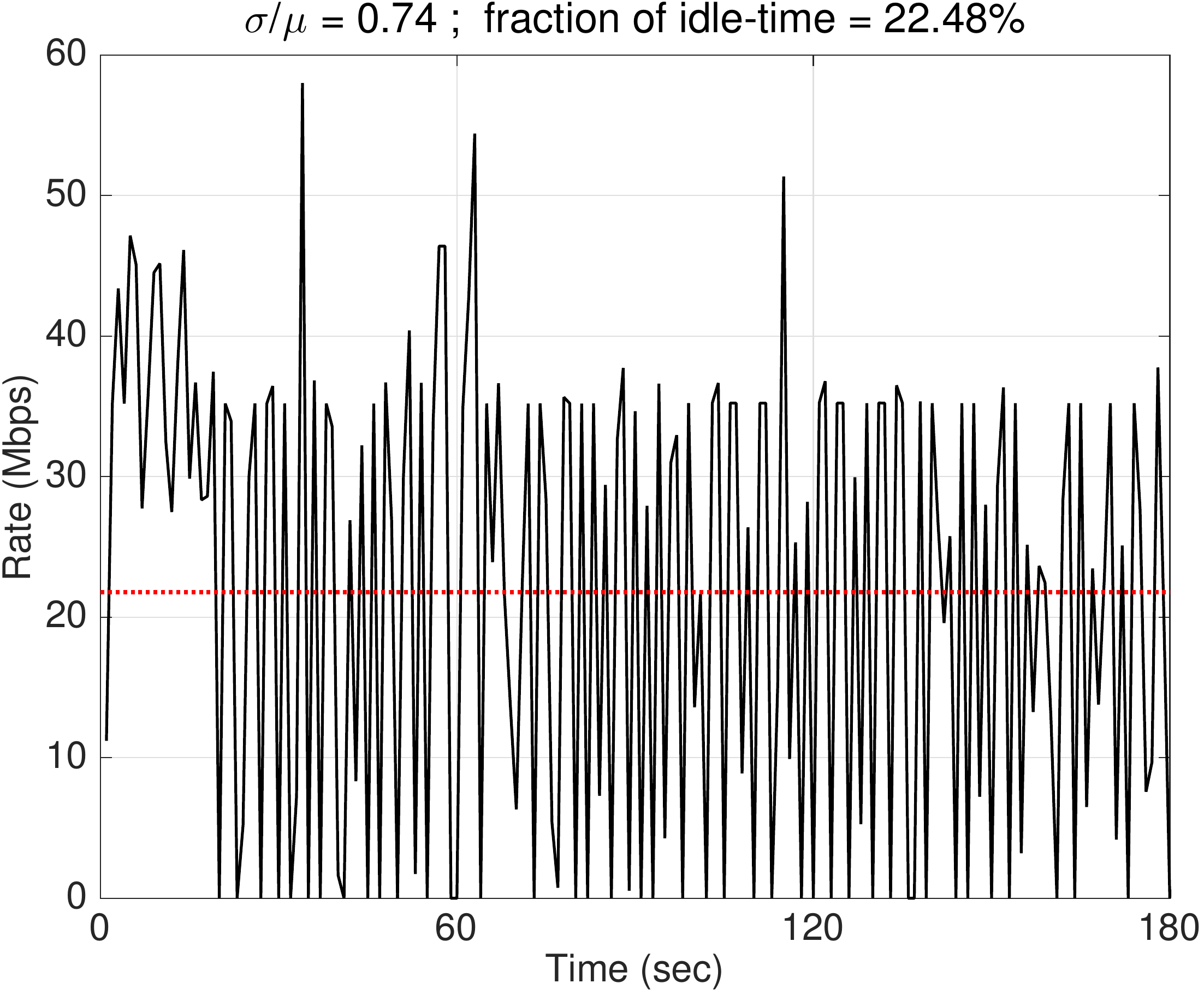}}\quad
				\label{fig:prof2160}
			}
		}
		\mbox{
		        \hspace{-4mm}
			\subfigure[Netflix (high resolution)]{
				{\includegraphics[width=0.24\textwidth]{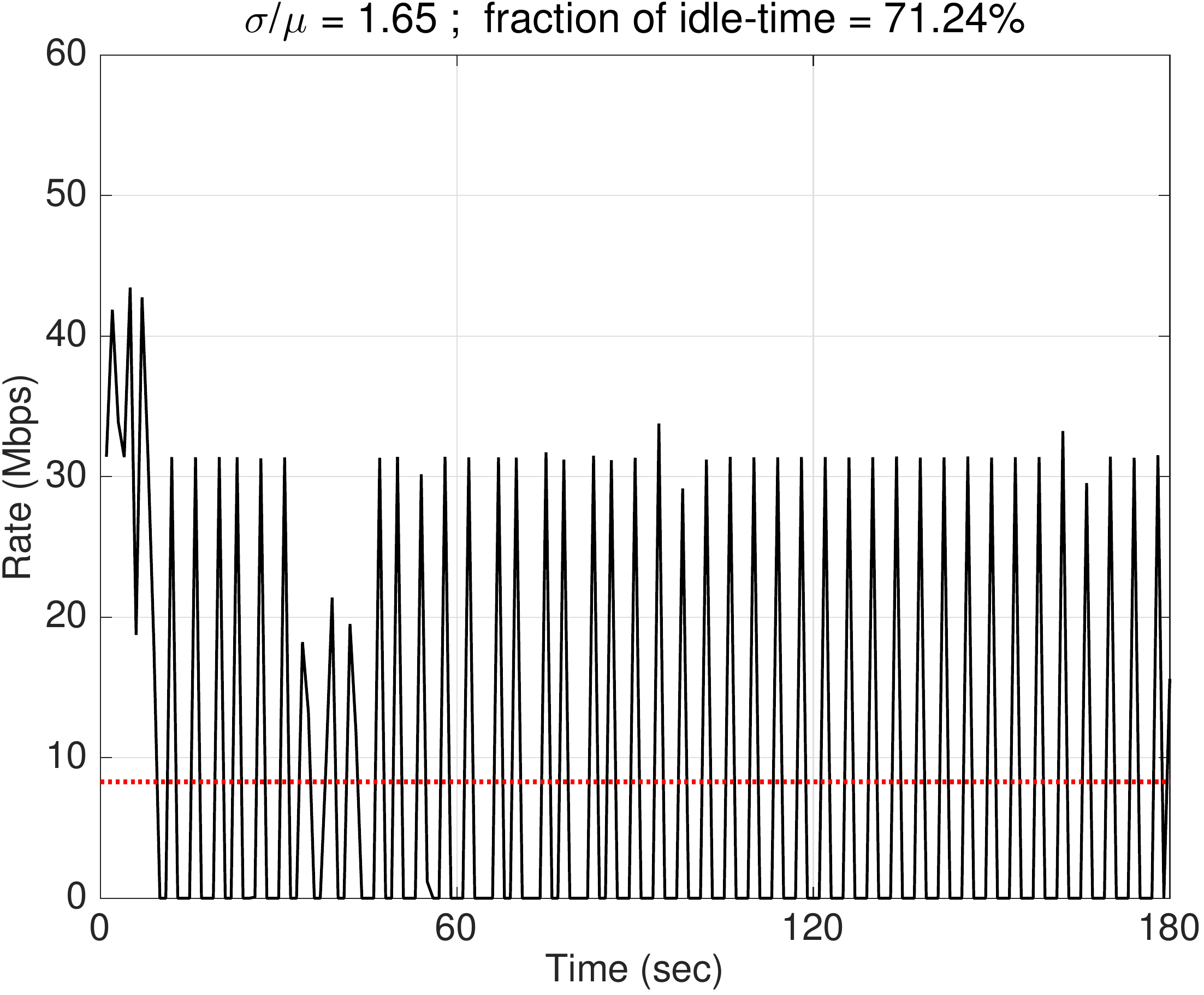}}\quad
				\label{fig:profnetflix}
			}
			\hspace{-4mm}
			\subfigure[Twitch (high resolution)]{
				{\includegraphics[width=0.24\textwidth]{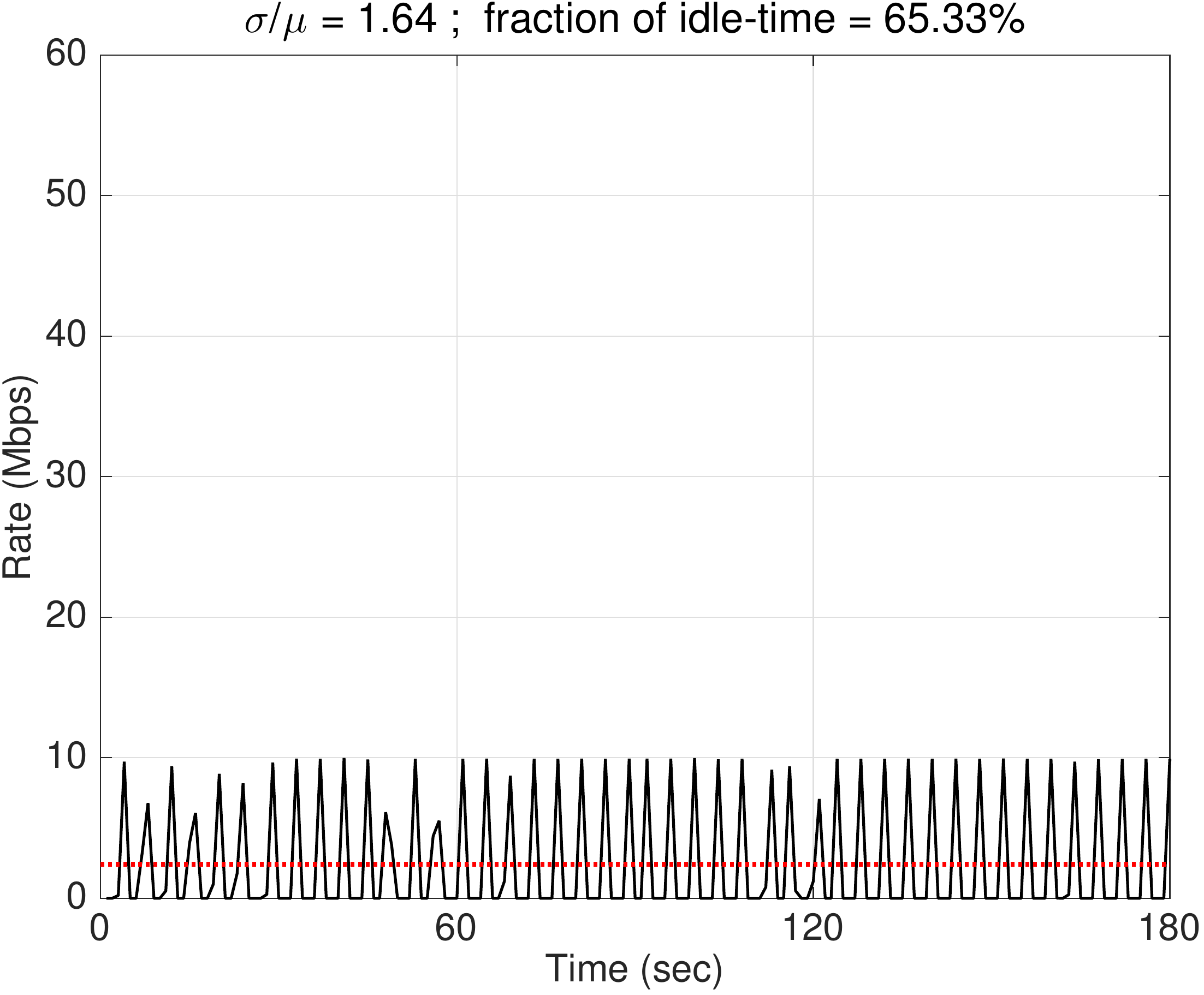}}\quad
				\label{fig:proftwitch}
			}
			\hspace{-4mm}
			\subfigure[Facebook app]{
				{\includegraphics[width=0.24\textwidth]{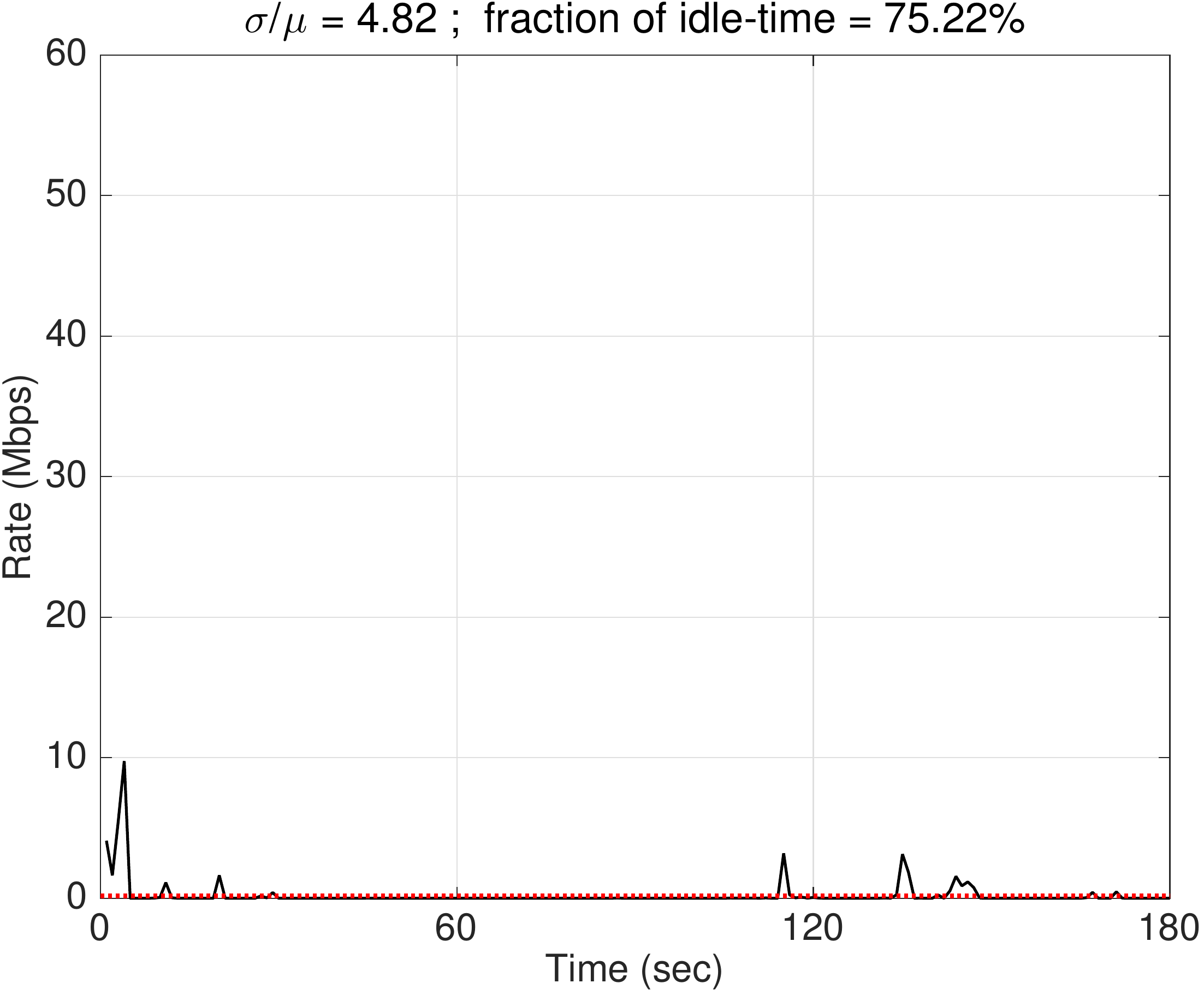}}\quad
				\label{fig:proffacebook}
			}
			\hspace{-4mm}
			\subfigure[Large download]{
				{\includegraphics[width=0.24\textwidth]{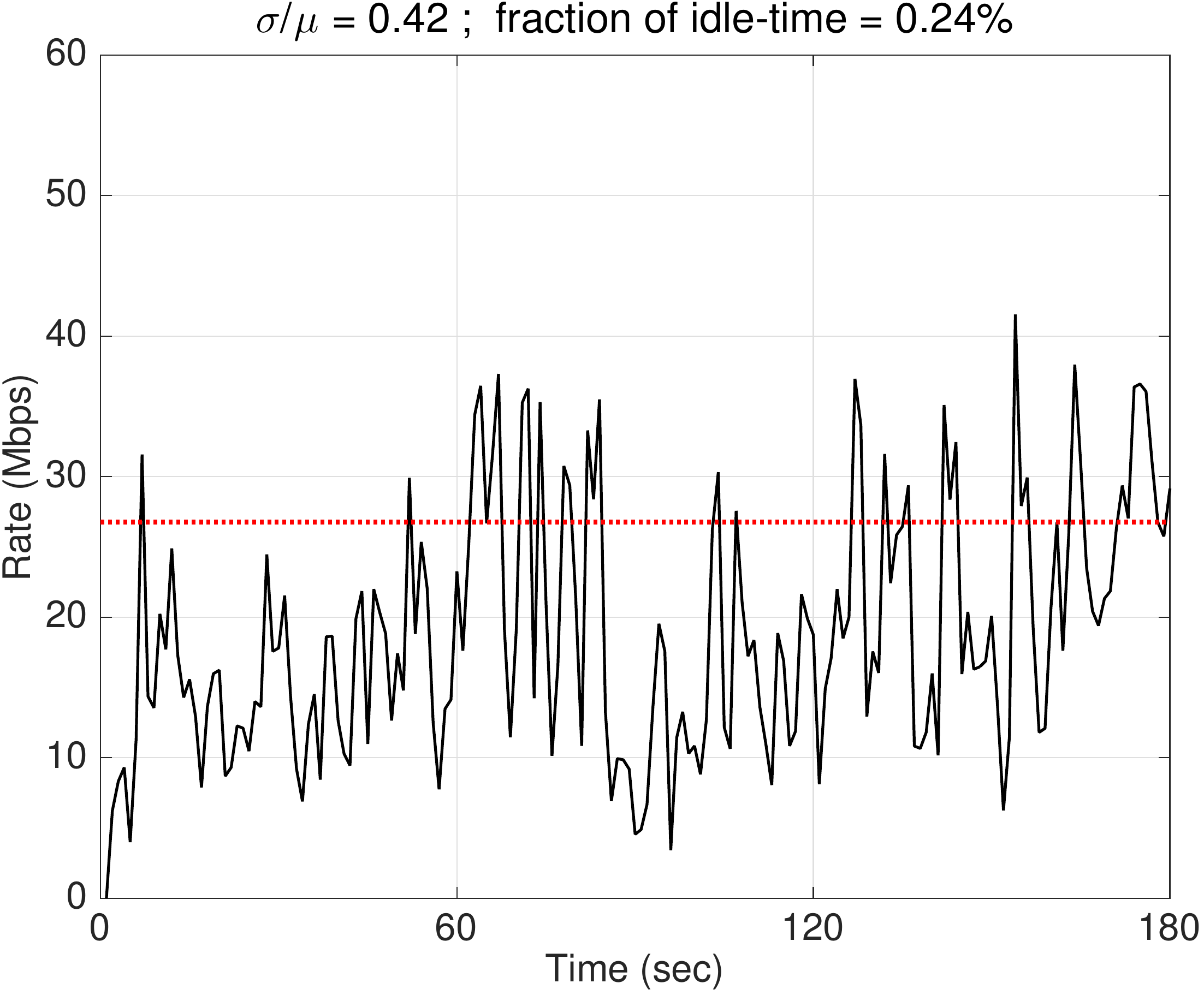}}\quad
				\label{fig:profDld}
			}
		}
		\caption{Traffic profiles.}
		\label{fig:prof}
	\end{center}
\end{figure*}

\subsection{Flow Table Management}\label{arch:rule}
We use a combination of proactive and reactive entries in a multi-table pipeline of the SDN switch, as shown in Fig.~\ref{fig:flowtable}. Reactive rules match on the 5-tuple, are of highest priority (Table 0), and are installed as a consequence of elephant flows detected by the packet inspection engine. They automatically time out upon a minute of inactivity, so as to reduce TCAM usage. The reactive flow entries achieve two objectives: to stop mirroring of long-flow packets to the software inspection engine, and to provide flow-level telemetry for the individual (potentially video) long-flows. The action corresponding to a match in the reactive table sends the flow to its appropriate entry in the {\em group table}, which identifies the content provider (Youtube, Netflix, etc.). The content provider for the flow is identified by searching for the server IP address in the most recent captured DNS suffixes (e.g. \verb|googlevideo.com| or \verb|nflxvideo.com|) that are stored in a time-series database by the software inspection engine. We note that if a video stream from a new DNS suffix is detected (e.g. \verb|ttnvw.net|) then a new group entry (for Twitch in this example) will be created dynamically. Our design not only makes the system adaptive to new video content providers, but also allows us to track aggregate video volumes for each video content provider.

Proactive (Table 1) entries are statically pushed by the controller so that all TCP (proto=6) and UDP (proto=17) packets received from the content provider, that have not already matched a elephant flow (Table 0), are forwarded (on port 2) and mirrored (on port 3) to the software inspection engine. Note that this includes DNS reply packets that contain the domain name of the video content provider and the video server IP address. All other packets are sent to Table 2, where the default action is to cross-connect the input and output ports (without any mirroring). We again emphasize that no data packets are sent to the controller, minimizing controller load, reducing packet-forwarding latency, and immunizing against controller failures.

\subsection{Packet Inspection Engine}\label{arch:anal}
The packet inspection engine keeps track of new flows, including 5-tuple information, duration, and volume, using efficient in-memory data structures. If a flow is active for more than a threshold volume, it is deemed as a elephant flow, and the engine informs the Broker which then makes a RESTful API call to the SDN controller to insert the reactive flow-table entry into the switch. This suppresses data-plane traffic for this flow from being mirrored to the inspection engine (as described in \S\ref{arch:rule}), and also triggers telemetry for that flow, as described in \S\ref{arch:algo}.

The other responsibility of the packet inspection engine is detection of DNS A-type replies, upon which it extracts the domain name and server IP addresses, and sends these via JSON to the broker, which writes it into a time-series DNS database. This database is used to associate a video stream to its content provider.  

\subsection{Telemetry Algorithm}\label{arch:algo}
Our data broker queries per-flow statistics (counters and timers), stores them in a time-series database, and exposes them to the user interface via appropriate RESTful APIs. The telemetry collects per-flow (fine grain) and per-group (coarse grain) usage statistics using the Stats collector module of our SDN application.

\textbf{Algorithm:} Recollect that our packet inspection engine identifies all elephant flows, which may include a mixture of video streams and elephant downloads, and suppresses their packets from being mirrored. We now develop an algorithm to distinguish video streams (from elephant transfers), and identify their content providers and resolutions. At a high level, the algorithm: (a) computes attributes of a given flow, which are then fed into an intelligent classifier (discussed in \S\ref{arch:ML}) to distinguish video streams from elephant transfers, (b) queries DNS database using the flow's client/server IP address to associate the video stream with its content provider and (c) estimates the resolution of the video stream (i.e. Low, Medium, High, Ultra-high).

\textbf{Usage Collection and Storage:} We collect flow counters per content provider (group table) and per video stream (reactive flow table). While the number of entries in the group table is generally small and fixed, the number of reactive flow entries can vary significantly with time. Polling the latter when the number of entries is large can result in a multi-part reply -- for example the Noviflow switch breaks the response into chunks of 2500 flows each -- putting considerable strain on the agent in the switch and affecting timeliness of the results. Prior work such as \cite{PayLess14} has explored the trade-off between accuracy, timeliness, and network overhead of polling switch entries, though their work is evaluated only in mininet emulation; in this work we take a relatively simplistic approach, whereby we tune the polling frequency depending on the number of entries. Consequently, when the number of reactive flows is less than 2500, we poll the counters every second, and the frequency reduces to once every 4 seconds when the number of entries increases to 10,000. The flow/group-level counters are stored in a time-series Flow DB, as shown in Fig.~\ref{fig:broker}, and are exposed periodically using a JSON-formatted message to the machine learning algorithm described next.

\subsection{Classification using Machine Learning}\label{arch:ML}
We develop a machine learning technique to determine if traffic pertaining to a flow is streaming video or not (the ``video identifier''), and if so, to determine the stream resolutions (the ``resolution classifier''). Our objective is to achieve video identification and classification in real-time with accuracy comparable to or better than computationally expensive techniques that require inspection of all traffic. 


\subsubsection{Attributes}
Attributes selection is of paramount importance for training of classifiers, given that these should be predictive to correctly identify/classify video streams.
To motivate our attributes selection and have an insight into behaviour of various flows, we plot in Fig.~\ref{fig:prof} the traffic pattern we have observed for various video streams including Youtube, Netflix and Twitch (at different resolutions: low, medium, high and ultra-high definition) and other elephant flows including Facebook application and large download (i.e. a representative of bulk transfer or GoogleDrive/dropbox cloud storage sychronization) during the first three minutes of their activity. 
It can be seen that due to buffering that accompanies video streaming, the {\em idle-time} characteristic (i.e. fraction of time that no data is exchanged) of video flows in Figs.~\ref{fig:prof144}-\ref{fig:proftwitch} is quite distinctive compared to the large download flow in Fig.~\ref{fig:profDld}). We also note that the {\em average rate} (shown by dotted red lines) of the Youtube 2160p (4k ultra-high definition video) in Fig.~\ref{fig:prof2160} is much higher than that of other video resolutions (shown in Figs.~\ref{fig:prof144}-\ref{fig:prof720} and ~\ref{fig:profnetflix}-\ref{fig:proftwitch}) but comparable to the large download in Fig.~\ref{fig:profDld}. In addition to idle-time and average rate, the {\em burstiness} characteristic of each flow is also distinctive -- the low resolution video and the large download exhibit the most and the least bursty patterns respectively, among these representative profiles shown in Fig.~\ref{fig:prof}. Based on these visual observations, we believe that idle-time, average rate and burstiness are collectively needed to identify and classify video flows. For example, the Facebook application flow shown in Fig.~\ref{fig:proffacebook} exhibits similar characteristics  of video streams (shown in Figures~\ref{fig:prof480}-\ref{fig:prof720}) in terms of idle-time and burstiness, but its rate is far below those of video streams.

The average rate and fraction of idle-time for a flow can be computed over a moving window (of say one minute). Burstiness of flow traffic can be computed in various ways \cite{InternetBurst05}, and it has been noted (particularly in the characterization of long-range dependent traffic) that it should be measured at multiple time-scales. We therefore compute the coefficient of variance (i.e. the ratio of the standard deviation to the mean, CV = $\sigma/\mu$) of our streams at time-granularities of 1-, 2-, 4-, 8- and 16-seconds, giving us $\sigma_1/\mu$, $\sigma_2/\mu$, $\sigma_4/\mu$, $\sigma_8/\mu$, and $\sigma_{16}/\mu$. These burstiness measures, in addition to idle-time and average rate $\mu$ of each flow, are input as attributes to our classifiers. Note that for a new flow, we may have only a subset of burstiness attributes at the begininng, as computing $\sigma_{16}$ would require collection of data for at least a minute. A flow that commenced only 20 seconds ago would only be able to yield $\sigma_1/\mu$, $\sigma_2/\mu$ and $\sigma_4/\mu$ since we have less than 4 data points at time scales of 8-second and 16-second.

\subsubsection{Identification/Classification}
As mentioned earlier, our iTeleScope data broker employs two classifiers namely the video identifier (to indicate if the flow is a streaming video or not) and the resolution classifier (to determine the resolution of video during playback). The identifier and classifier are invoked every 16 seconds to dynamically capture profile changes (e.g. video stream rate adaptation) -- initial invocation may have access to only five attributes (idle-time, $\mu$, $\sigma_1/\mu$, $\sigma_2/\mu$, and $\sigma_4/\mu$), and subsequent invocations that have access to more (burstiness-related) attributes may change the classification, improving accuracy and/or identifying resolution changes. The training of the classifiers will be described in the next section.

\begin{figure}[!t]
	\centering
	\hspace{-5mm}
	\includegraphics[width=0.495\textwidth]{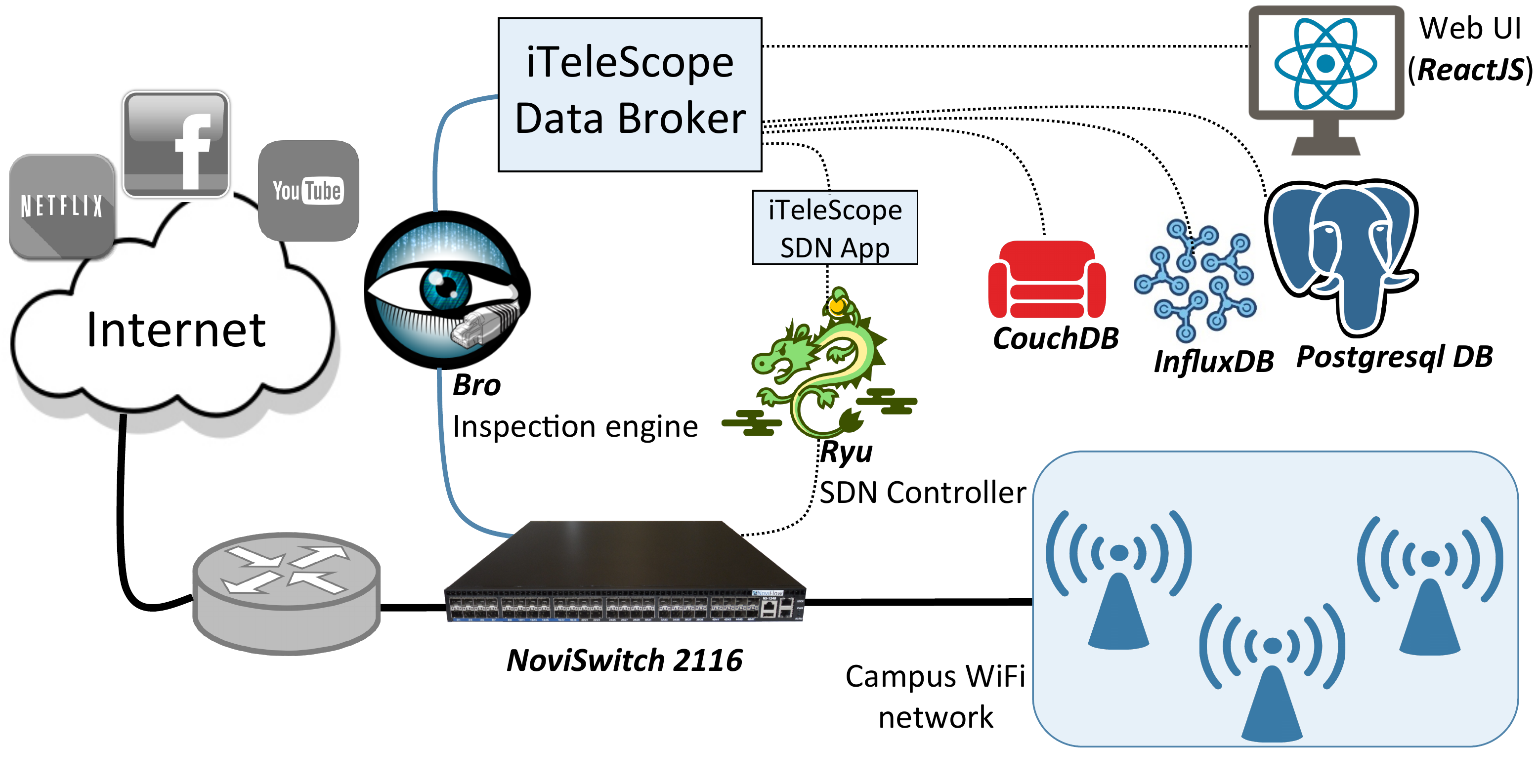}
	\caption{iTeleScope prototype}
	\label{fig:imp}
\end{figure}

\section{Prototype Implementation and Machine Training}\label{proto}
\subsection{Prototype}
We have implemented a fully functional near-production-grade iTeleScope system that identifies and classifies video streams in real-time at line-rate of up to 10 Gbps. For our system we have implemented an application on top of the Ryu SDN controller, augmented the Bro packet inspection engine for flow state management and event triggering, implemented various databases including InfluxDB, PostgreSQL, and CouchDB, and a web-GUI (in ReactJS) for interaction with our tool. Further, each of these components operates on a separate docker container or virtual machine in our cloud environment powered by the VMware Esxi 6.0 hypervisor. All VMs run Ubuntu server 14.04 LTS and are allocated to a four-core CPU, with 8 GB of memory and 32 GB disk space. Our system is currently managing three environments: (a) an SDN-enabled experimental lab network connected via WiFi access points (used for machine training in \S\ref{MLtraining}), (b) a point-to-point link over which an industrial scale Spirent traffic generator feeds traffic into our setup, and (c) a live campus dorm network link operating at 10 Gbps and serving several hundred real users. Our implemented design is depicted in Fig.~\ref{fig:imp}, and the interested reader can see the interface live via our publicly accessible website at: \verb|https://telescope.sdn.unsw.edu.au/|.

\begin{figure}[t!]
	\centering
	
	\mbox{\includegraphics[width=0.48\textwidth]{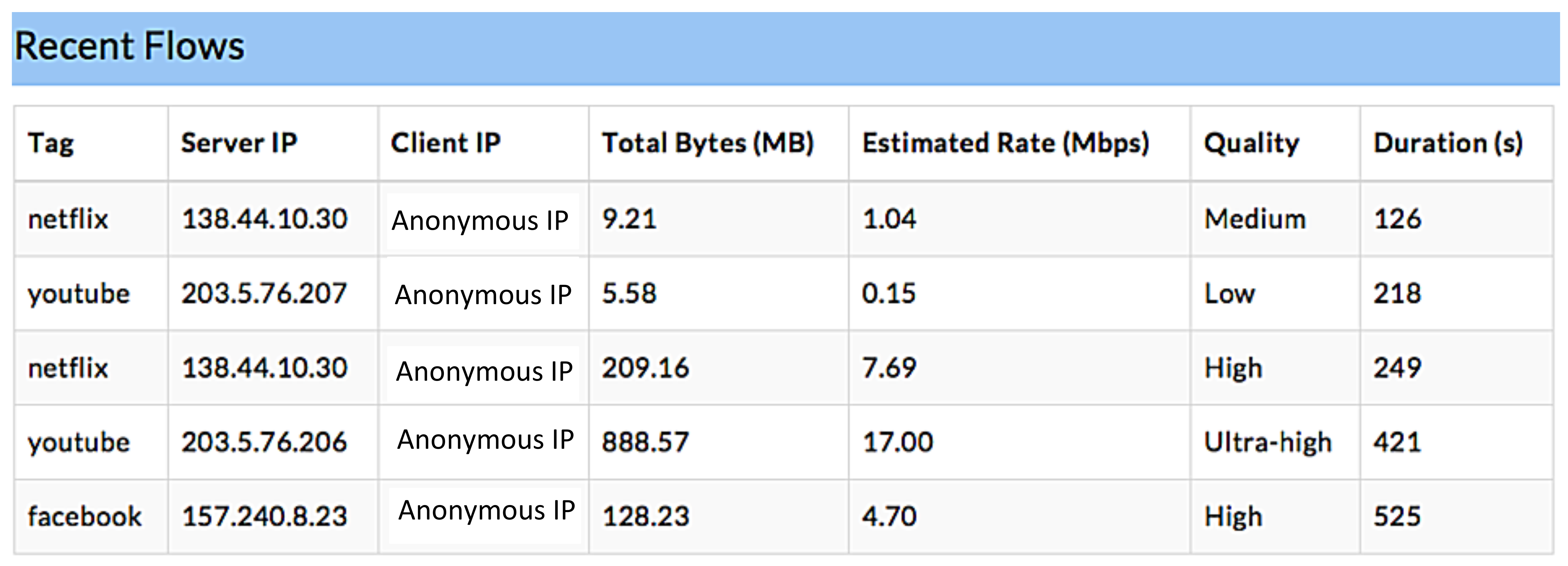}}
	\caption{Recent video flows.}
	\label{fig:campus:recentFlows}
\end{figure}

\begin{figure}[t!]
	\centering
	\mbox{\includegraphics[width=0.48\textwidth]{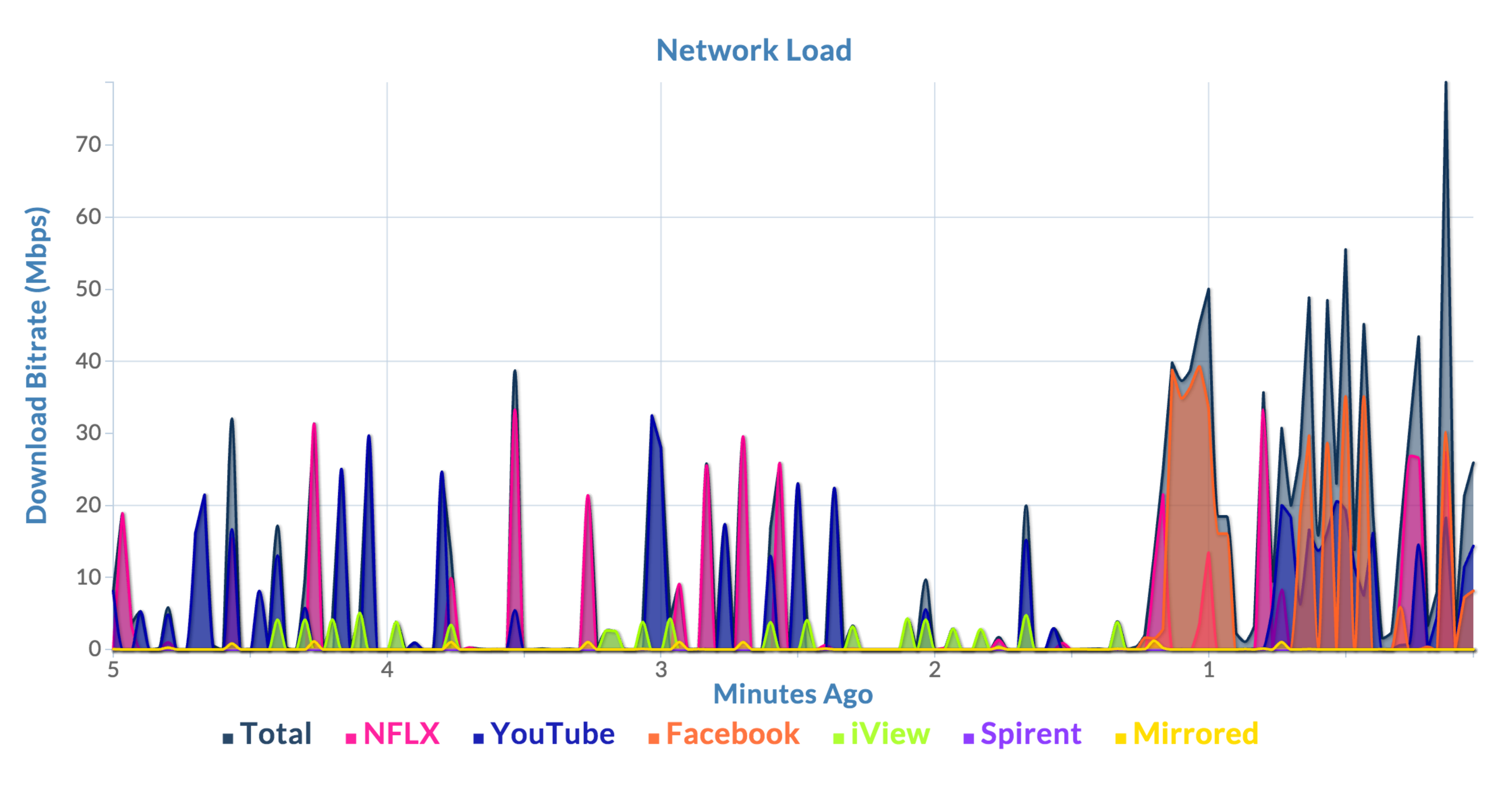}}
	\caption{Load of the network and video traffic.}
	\label{fig:campus:load}
\end{figure}

\textbf{SDN switch}: Our SDN switch is a fully Openflow 1.3 compliant NoviSwitch 2116, as shown in Fig.~\ref{fig:imp}. It provides 160 Gbps of throughput, tens of thousands of TCAM flow entries, and millions of exact-match flow-entries in DRAM, and we found it to amply cater to the requirements of this project.

\textbf{Packet Inspection Engine:} We use the Bro (v2.4.1) \cite{Bro} open-source tool for inspection of the mirror traffic. We wrote event-handlers in Bro that keep track of the flow duration and volume, and to trigger an API call to the data broker when a elephant flow is detected. Similarly, DNS replies are also parsed and the information passed to the data broker for recording into the time-series database.

\textbf{Data broker:} We used python to implement our data broker that receives the 5-tuple of elephant flows and DNS information from the Bro inspection engine, inserts/modifies flow/group entries, and collects statistical data from our SDN application via RESTful API. Flow and group stats collected from the SDN application are written into a time series InfluxDB. Flow level information is queried from InfluxDB periodically for processing by the intelligent classifier trained by the Weka tool \cite{Weka} using Weka's Python library wrapper interface (v0.3.9). The intelligent classifier identifies video flows, queries the DNS database to label video flows, calls RESTful APIs to modify flow entires' output group, and identifies video stream resolutions, as described in \S\ref{arch:rule}.

\textbf{SDN controller and application:} We used the Ryu (v4.0) Openflow controller for operating our system, and developed a Python based SDN application exposing northbound RESTful APIs to the data broker for inserting or modifying network rules and polling flow statistics. Successful RESTful API calls result in appropriate actions (e.g. network rules insertion, modification and counters collection) at the SDN switch serving the data-plane. 

\textbf{DataBases}: We employ three databases in our system to store flows usage statistics, DNS information, and system configurations. We use time-series InfluxDB (v1.0.0) to store periodic flow/group statistics as mentioned in \S\ref{arch:algo}. In the same InfluxDB we also store information of DNS A-type replies including the domain name and client/server IP addresses. An object relational database PostgreSQL (v9.6.3) is used to store the mapping between domain IP addresses, domain name suffixes and provider names. Lastly, we use a NoSQL CouchDB (v2.0.0) document-oriented database to store configurations of the SDN switch such as DPID and multi-table configs.
\begin{figure*}[t!]
	\begin{center}
		\mbox{
			\subfigure[Idle-time]{
				{\includegraphics[width=0.24\textwidth,height=.20\textwidth]{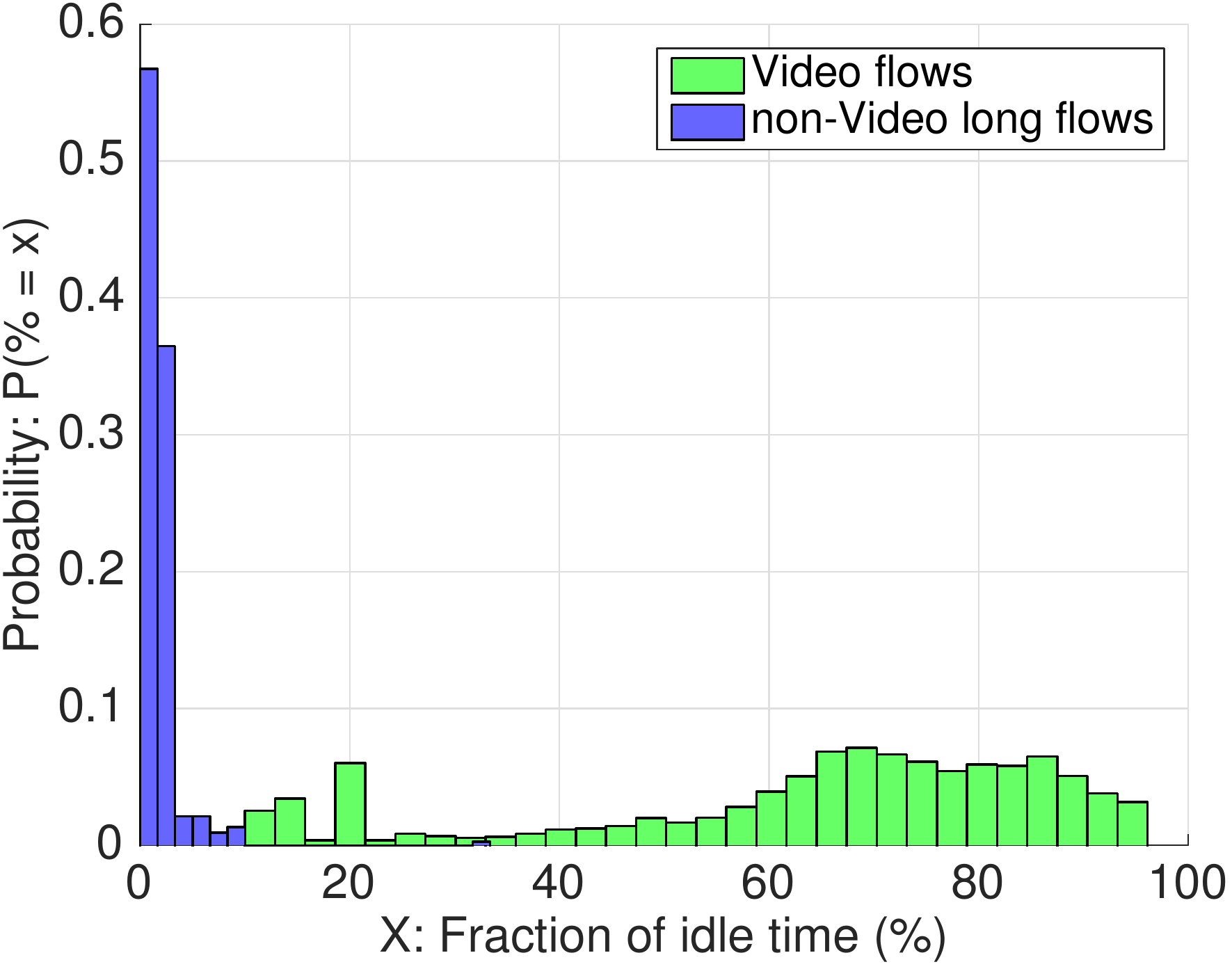}}\quad
				\label{fig:histIdle}
			}
			\hspace{-6mm}
			\subfigure[average rate]{
				{\includegraphics[width=0.24\textwidth,height=.20\textwidth]{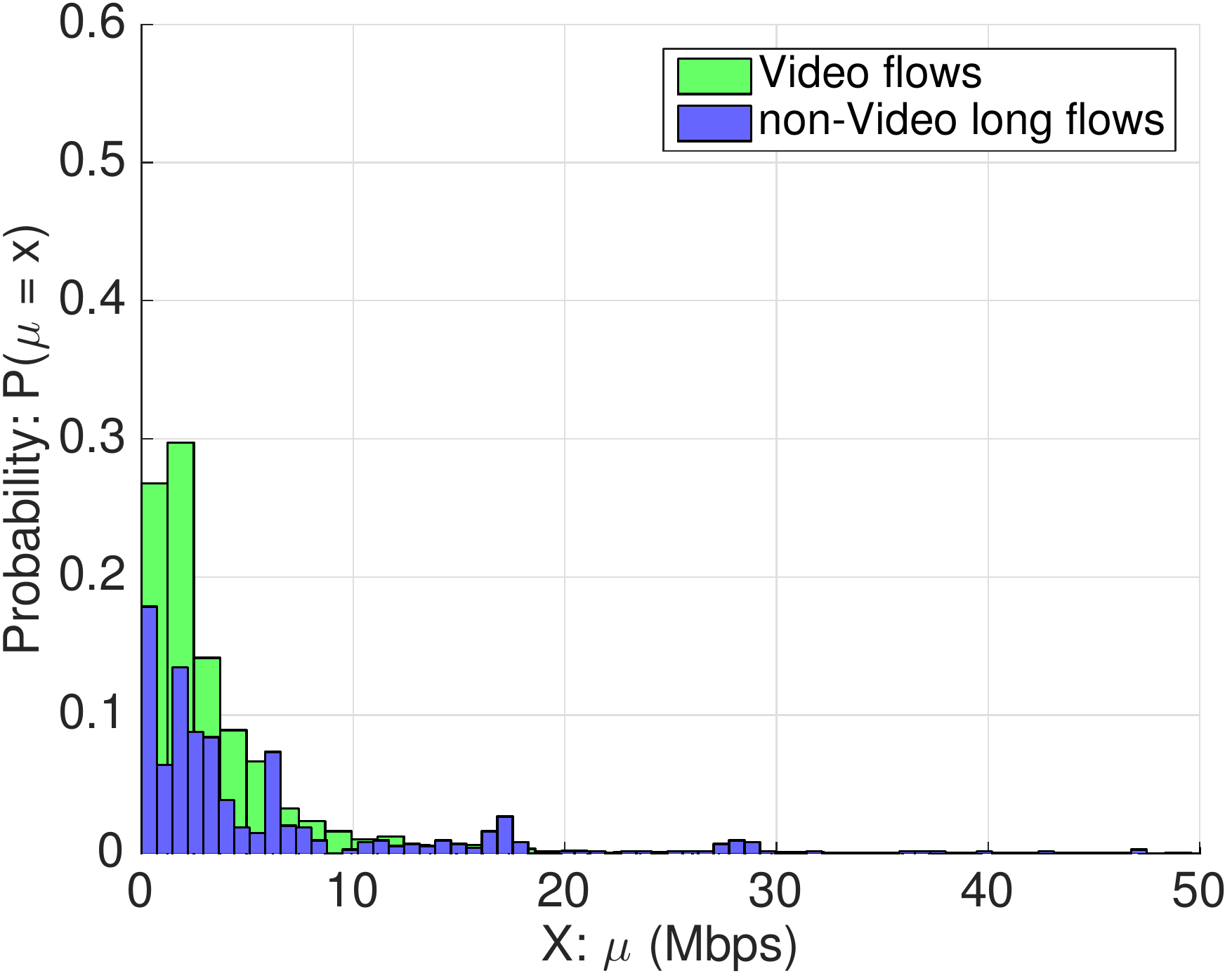}}\quad
				\label{fig:histRate}
			}
			\hspace{-6mm}
			\subfigure[burstiness at 1-second]{
				{\includegraphics[width=0.24\textwidth,height=.20\textwidth]{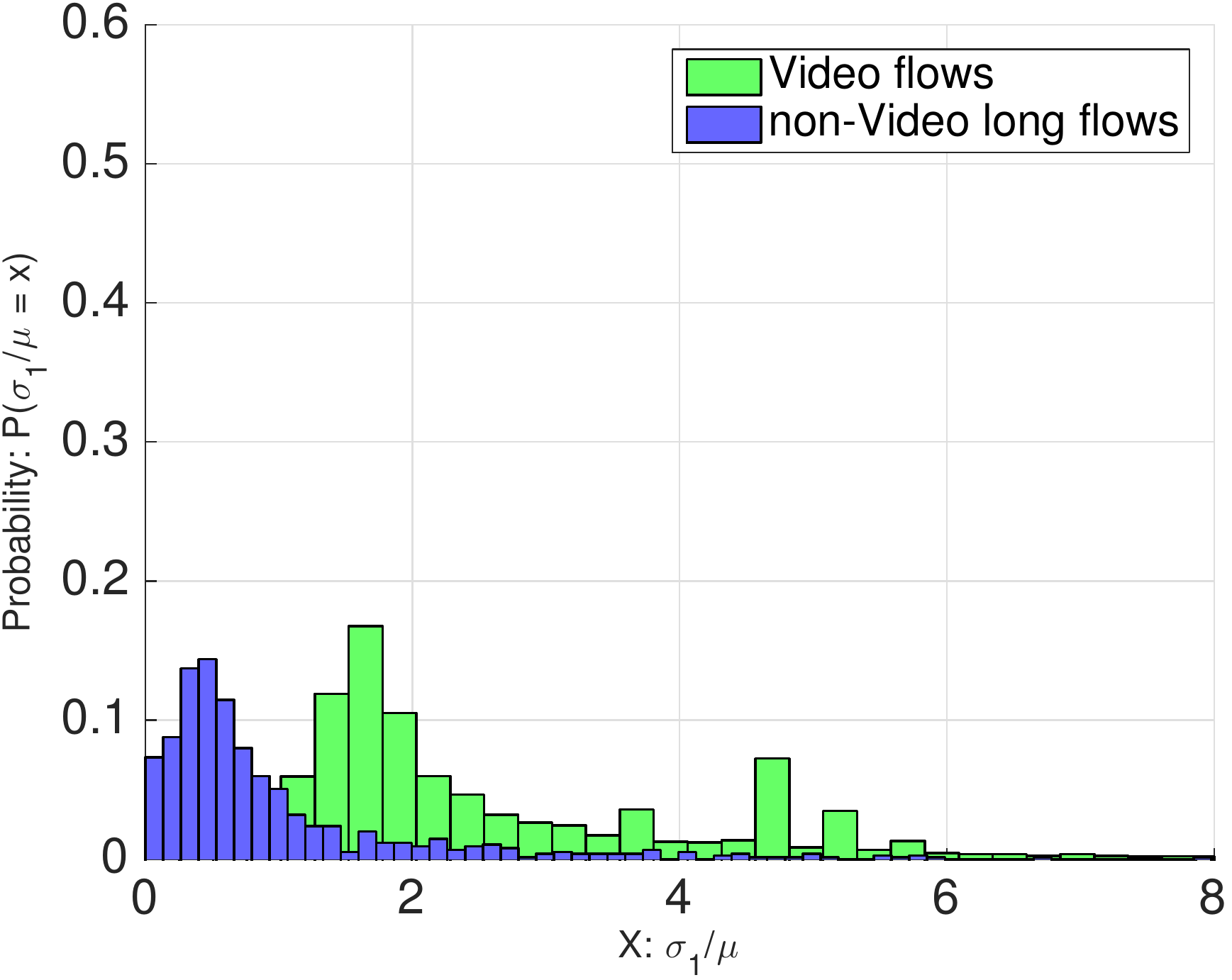}}\quad
				\label{fig:histSig1}
			}
		      \hspace{-6mm}
			\subfigure[burstiness at 2-second]{
				{\includegraphics[width=0.24\textwidth,height=.20\textwidth]{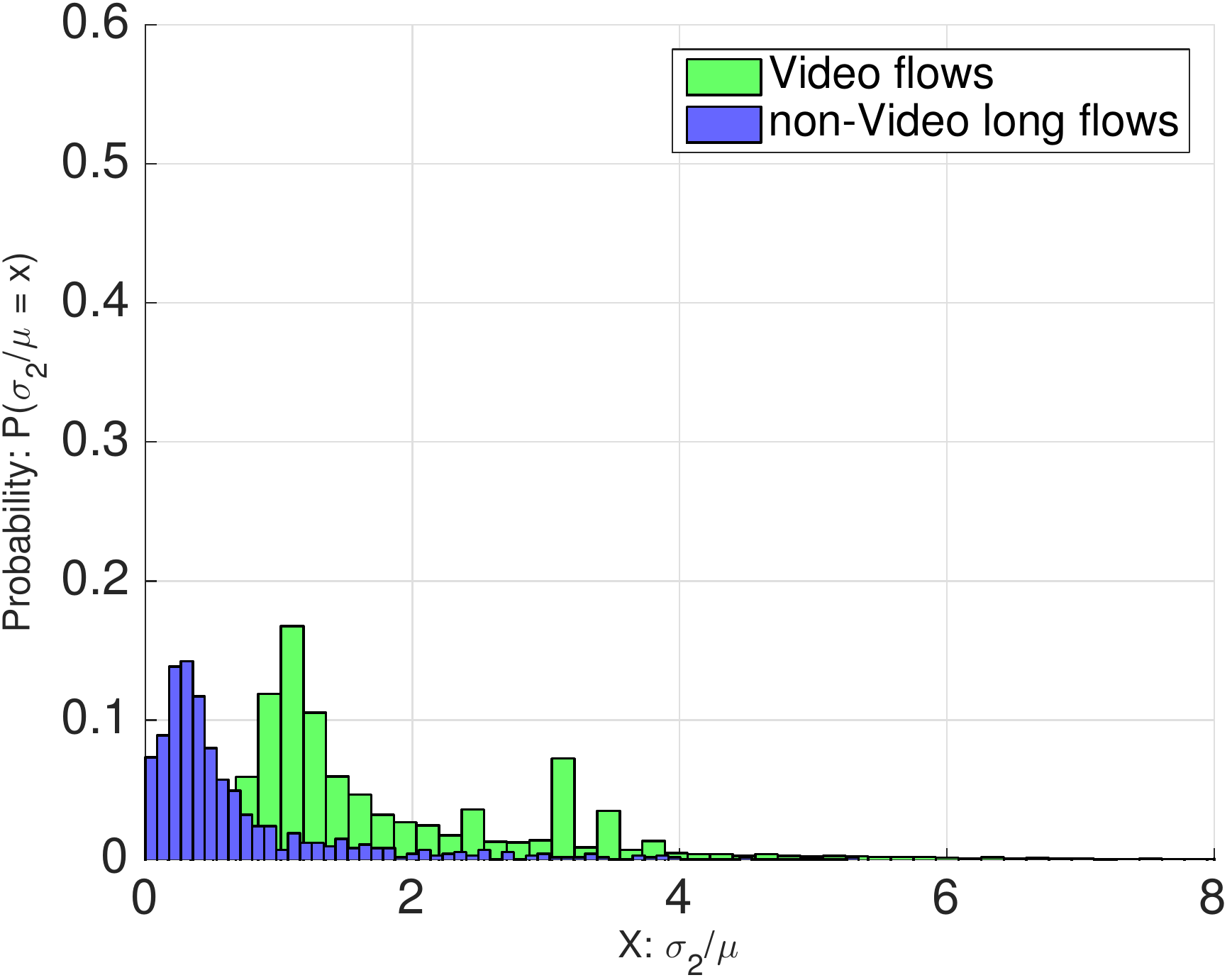}}\quad
				\label{fig:histSig2}
			}
		}
		\mbox{			
                 \hspace{-3mm}
			\subfigure[burstiness at 4-second]{
				{\includegraphics[width=0.32\textwidth,height=.20\textwidth]{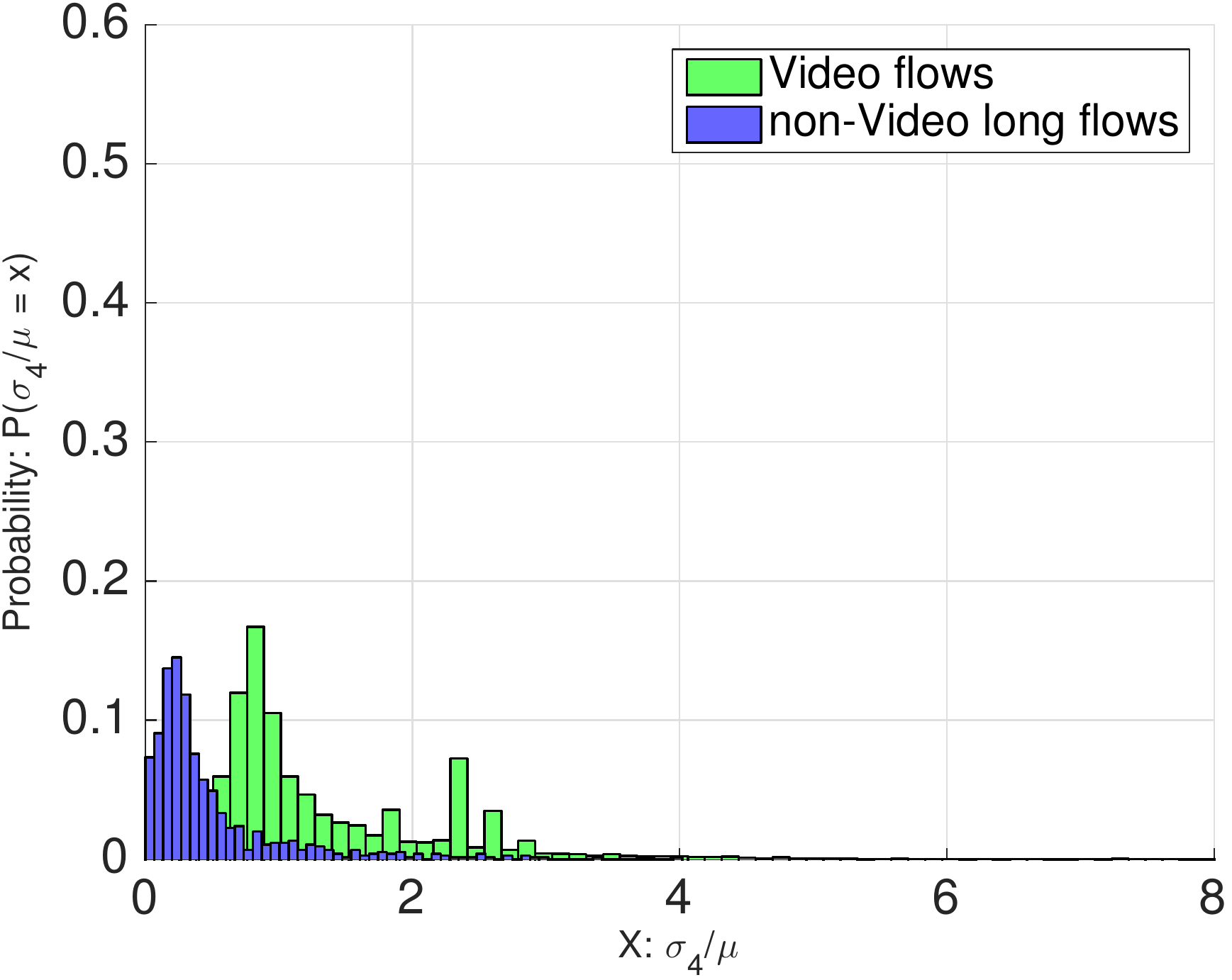}}\quad
				\label{fig:histSig4}
			}
			\hspace{-5mm}
			\subfigure[burstiness at 8-second]{
				{\includegraphics[width=0.32\textwidth,height=.20\textwidth]{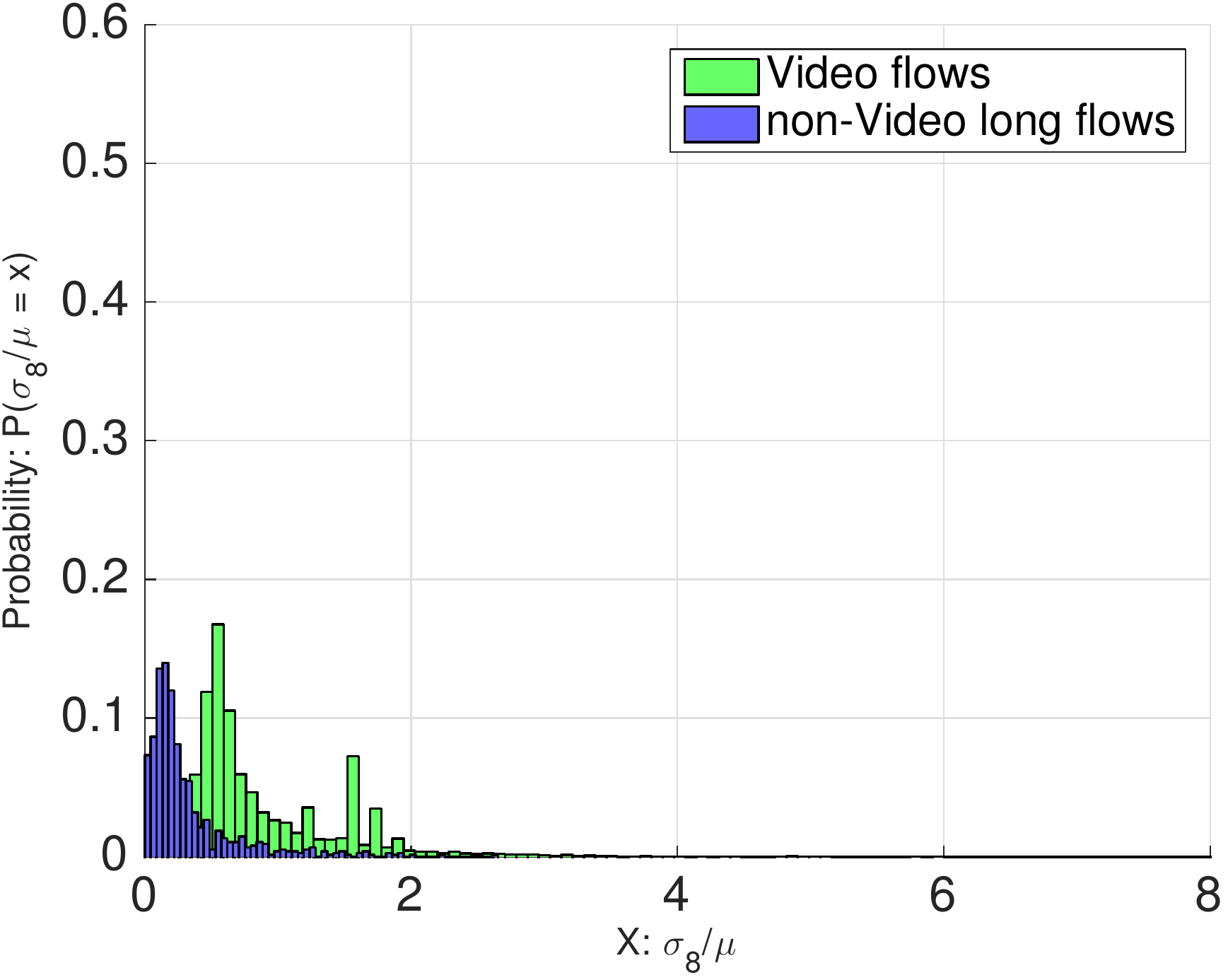}}\quad
				\label{fig:histSig8}
			}
			\hspace{-5mm}
			\subfigure[burstiness at 16-second]{
				{\includegraphics[width=0.32\textwidth,height=.20\textwidth]{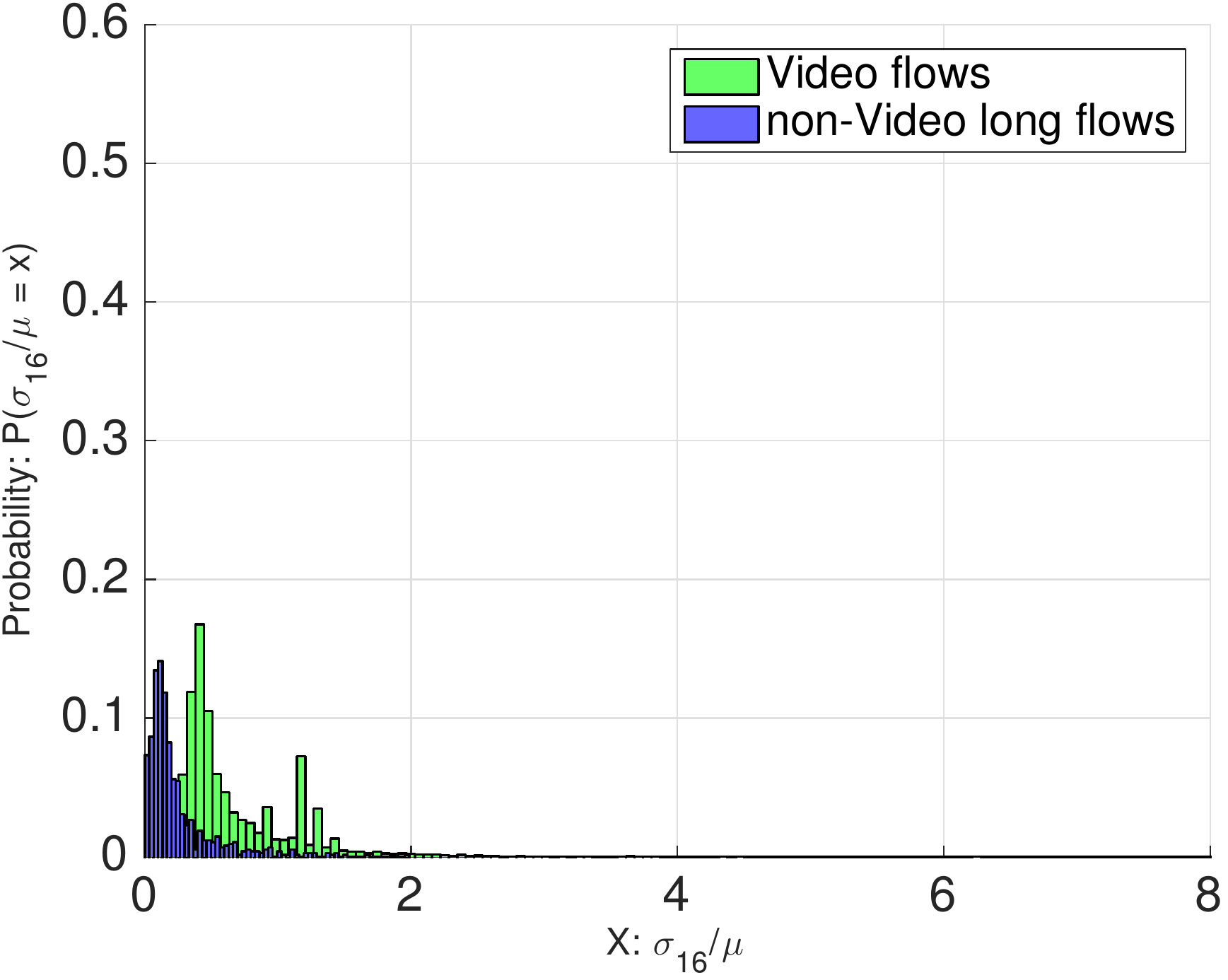}}\quad
				\label{fig:histSig16}
			}
		}				
		\vspace{-3mm}
		\caption{Histogram of idle-time, average rate and burstiness at various time scales (video vs. non-video). }
		\vspace{-3mm}
		\label{fig:hist}
	\end{center}
\vspace{-2mm}
\end{figure*}

\textbf{Web Interface}: We provide a front-end for network operators to visualize video streams in their network, implemented in ReactJS using Rubix template and D3 library. Snapshots are shown in Fig.~\ref{fig:campus:recentFlows} and~\ref{fig:campus:load}. The reader can see the live interface via a public accessible website at: \verb|https://telescope.sdn.unsw.edu.au|. The interface shows aggregated video consumption statistics by different content providers over the last one hour, one day and one week, the total number of elephant flows, and the most recent video streams.

\subsection{Machine Training}\label{MLtraining}
We now train our classifiers with datasets collected in our lab over our prototype. In order to have the ground truth for the training, we scripted the streaming of video from various providers (i.e. Youtube, Netflix, Youku, Facebook, Tencent) at various resolutions. The automation was done using APIs where possible, such as the Youtube Player API that allows videos to be streamed at specified resolution (i.e. low: 144p, 240p, 360p; medium: 480p, 720p; high: 1080p, 1440p; and ultra-high: 4K), and by launching a browser URL otherwise. We also scripted the generation of elephant flow traffic including large ISO file downloads and Google-Drive sync operations, and mice flows from dynamic webpage loads (e.g. Office 365, Facebook homepage, WhatsApp).

For the purpose of training, our scripts limit all flows (video and non-video) to 128 seconds (i.e. about two minutes), even though all chosen videos have total length in excess of 20 minutes. At the end of each two-minute activity, the script queries the InfluxDB to extract the flow profile (byte counts at 1-second time interval) to calculate attributes (as explained in \S\ref{arch:ML}). We then split the 128-second traffic profile into 8 sub-profiles (i.e. time intervals of [1,16]s, [1,32]s, [1,48]s, [1,64]s, [17, 80]s, [33, 96]s, [48, 112]s, and [65, 128]s). The script lastly computes the attributes for each of the sub-profiles. We note that short sub-profiles (e.g. [1,16]s) will have incomplete attributes such as $\sigma_8/\mu$ and $\sigma_{16}/\mu$. We have run our script for two weeks and collected a total of 28,543 labeled training instances for elephant flows (video and non-video) of which 10,416 instances were labeled for various video resolutions.

\subsubsection{Attribute Profiles}
We briefly describe the profiles of the attributes (idle-time, average rate, and burstiness at various time-scales) collected from our dataset. Fig.~\ref{fig:hist} shows the histogram of these attributes used by the video identifier machine. The difference between video and non-video long flows is visually apparent: for example, Fig.~\ref{fig:histIdle} shows that non-video flows have very low idle time fraction (centered at about 1\% with minor deviations), whereas video traffic idle-time fraction is widely spread between 20\% and 95\%. The video and non-video streams are not so distinct in their distribution of average rate (Fig.~\ref{fig:histRate}); however, they do have different burstiness behaviors at various time-scales, as seen in Figs.~\ref{fig:histSig1}-\ref{fig:histSig16}. 

\begin{figure*}[t!]
	\begin{center}
		\mbox{
			\subfigure[idle-time]{
				{\includegraphics[width=0.24\textwidth,height=.20\textwidth]{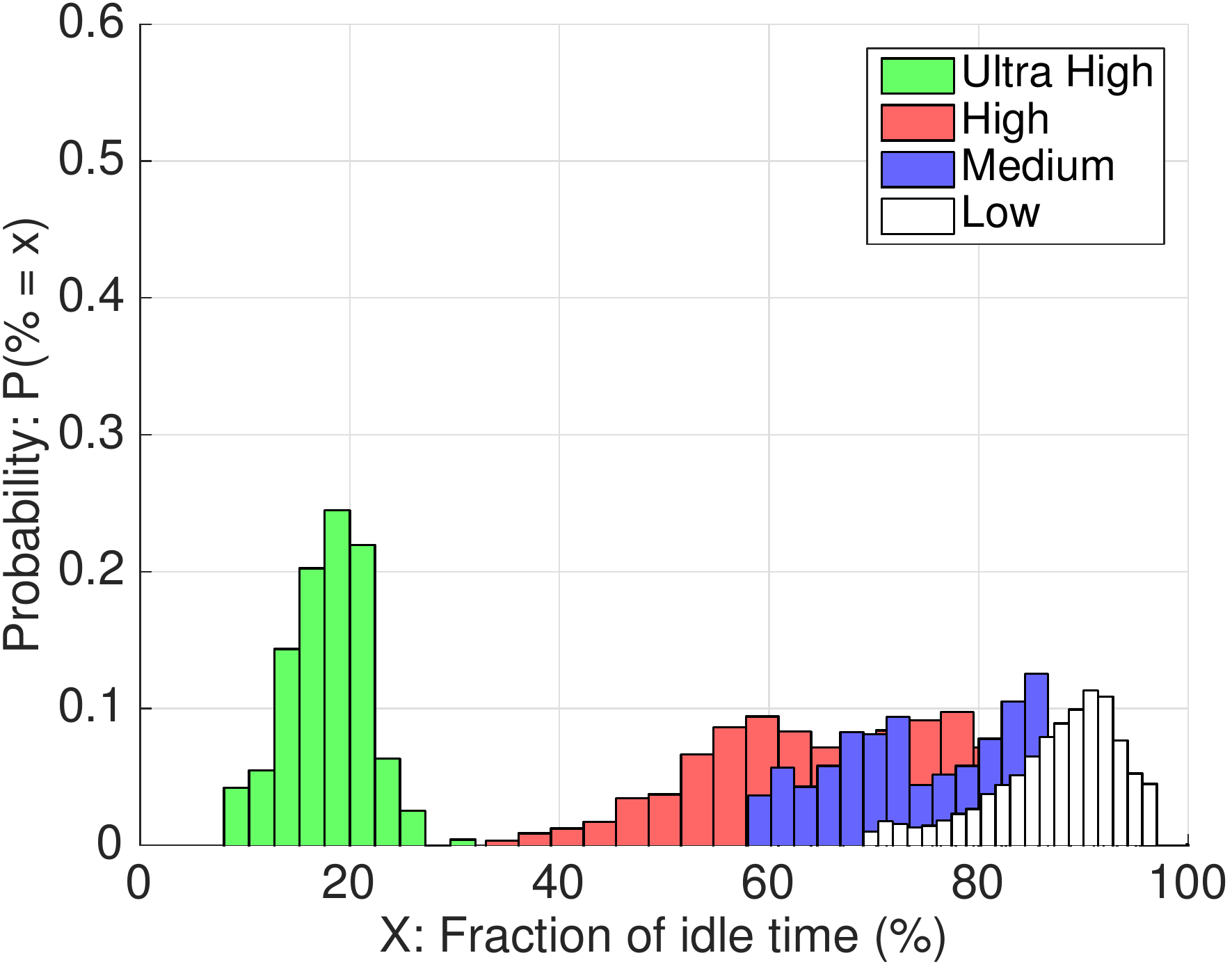}}\quad
				\label{fig:histIdleVid}
			}
			\hspace{-6mm}
			\subfigure[average rate]{
				{\includegraphics[width=0.24\textwidth,height=.20\textwidth]{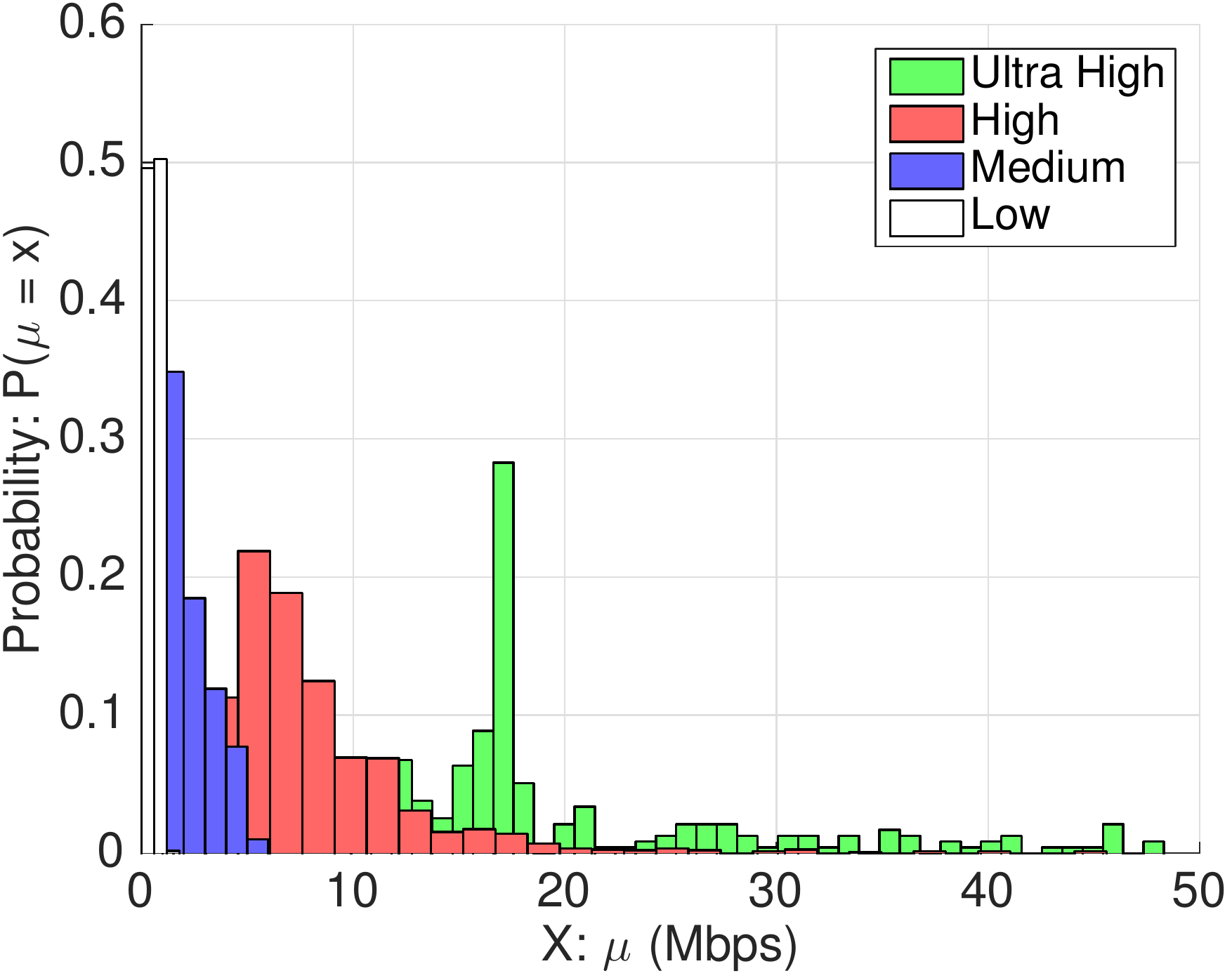}}\quad
				\label{fig:histRateVid}
			}
			\hspace{-6mm}
			\subfigure[burstiness at 1-second]{
				{\includegraphics[width=0.24\textwidth,height=.20\textwidth]{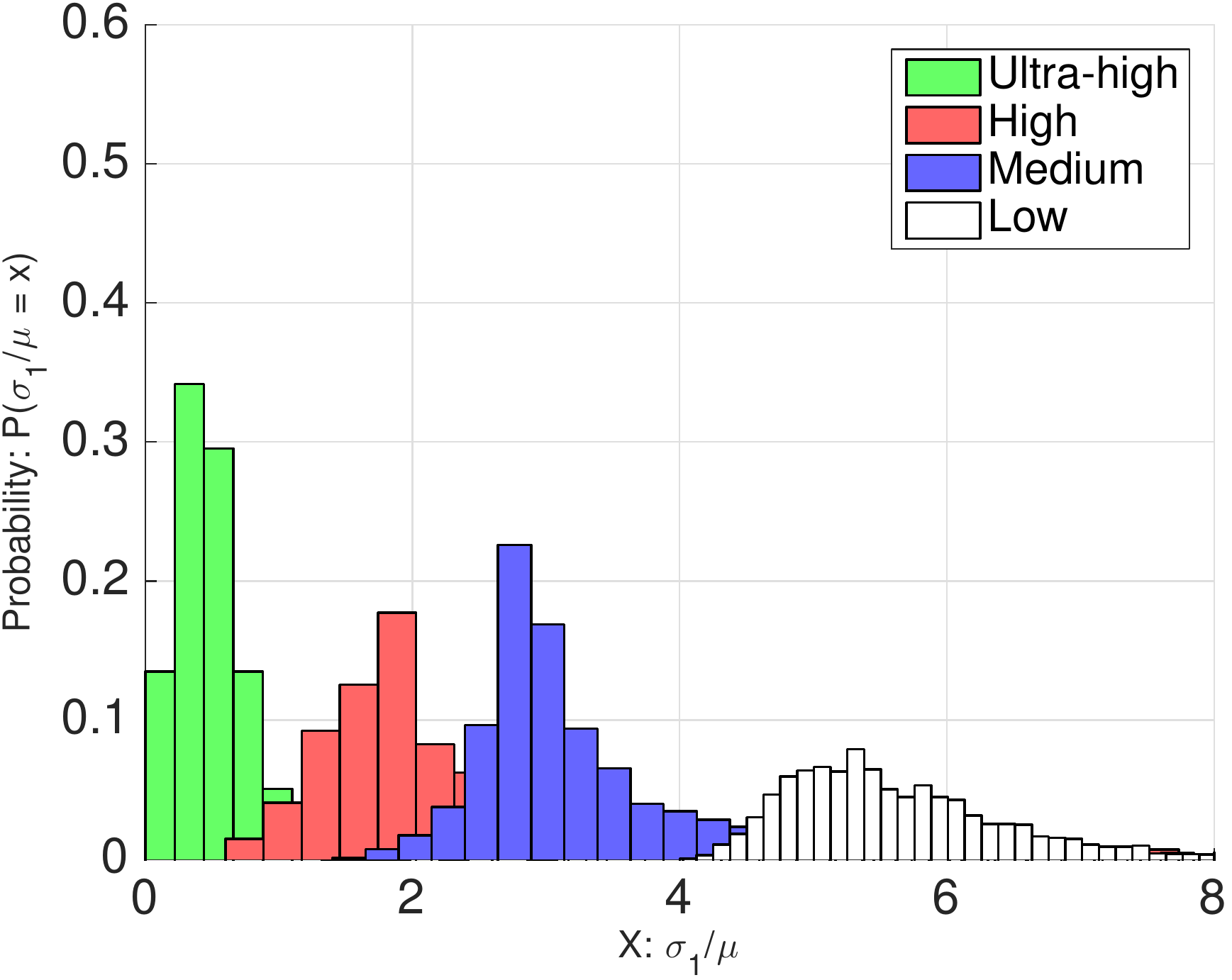}}\quad
				\label{fig:histSig1Vid}
			}
			\hspace{-6mm}
			\subfigure[burstiness at 2-second]{
				{\includegraphics[width=0.24\textwidth,height=.20\textwidth]{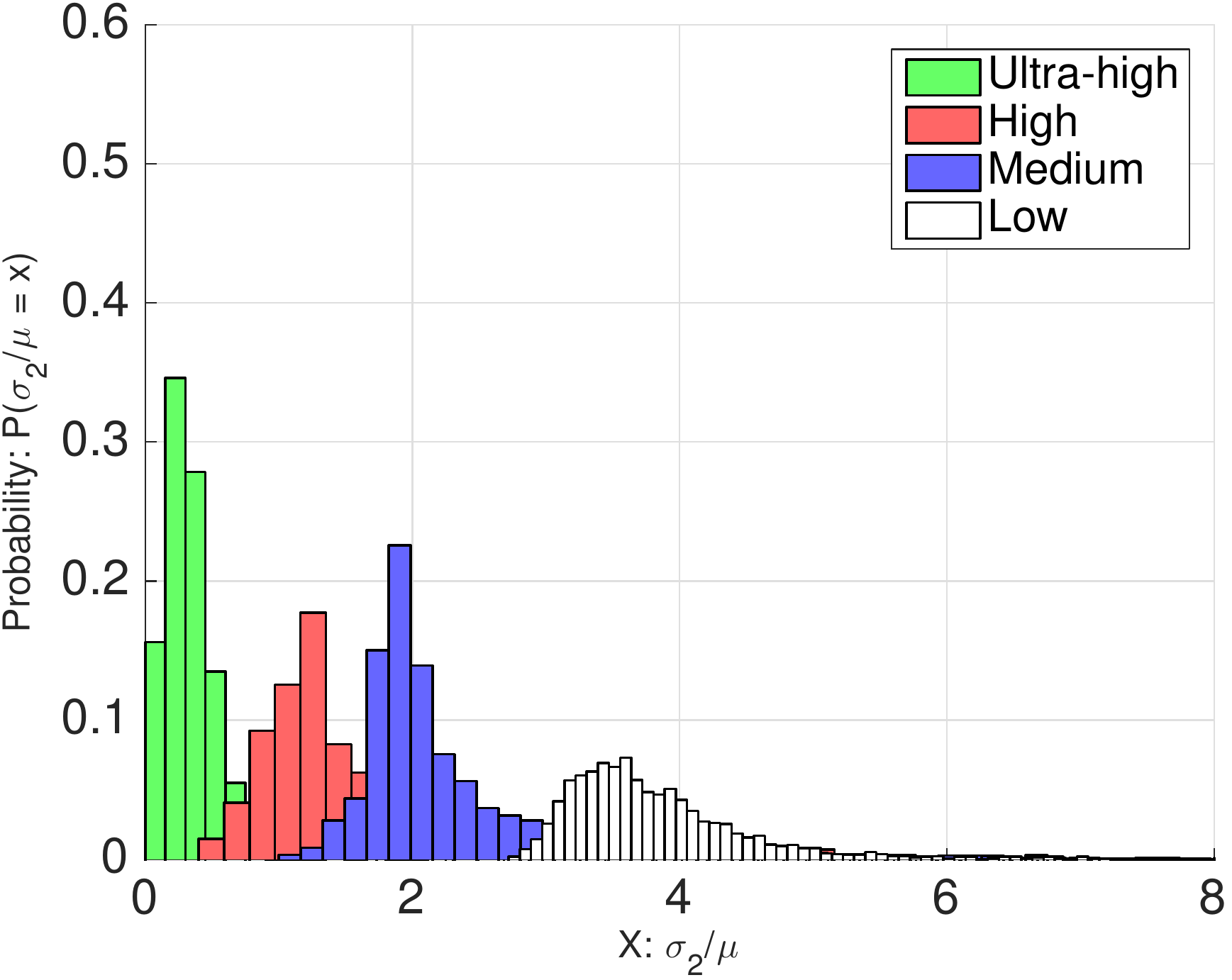}}\quad
				\label{fig:histSig2Vid}
			}
		}
		\mbox{			
			\hspace{-3mm}
			\subfigure[burstiness at 4-second]{
				{\includegraphics[width=0.32\textwidth,height=.20\textwidth]{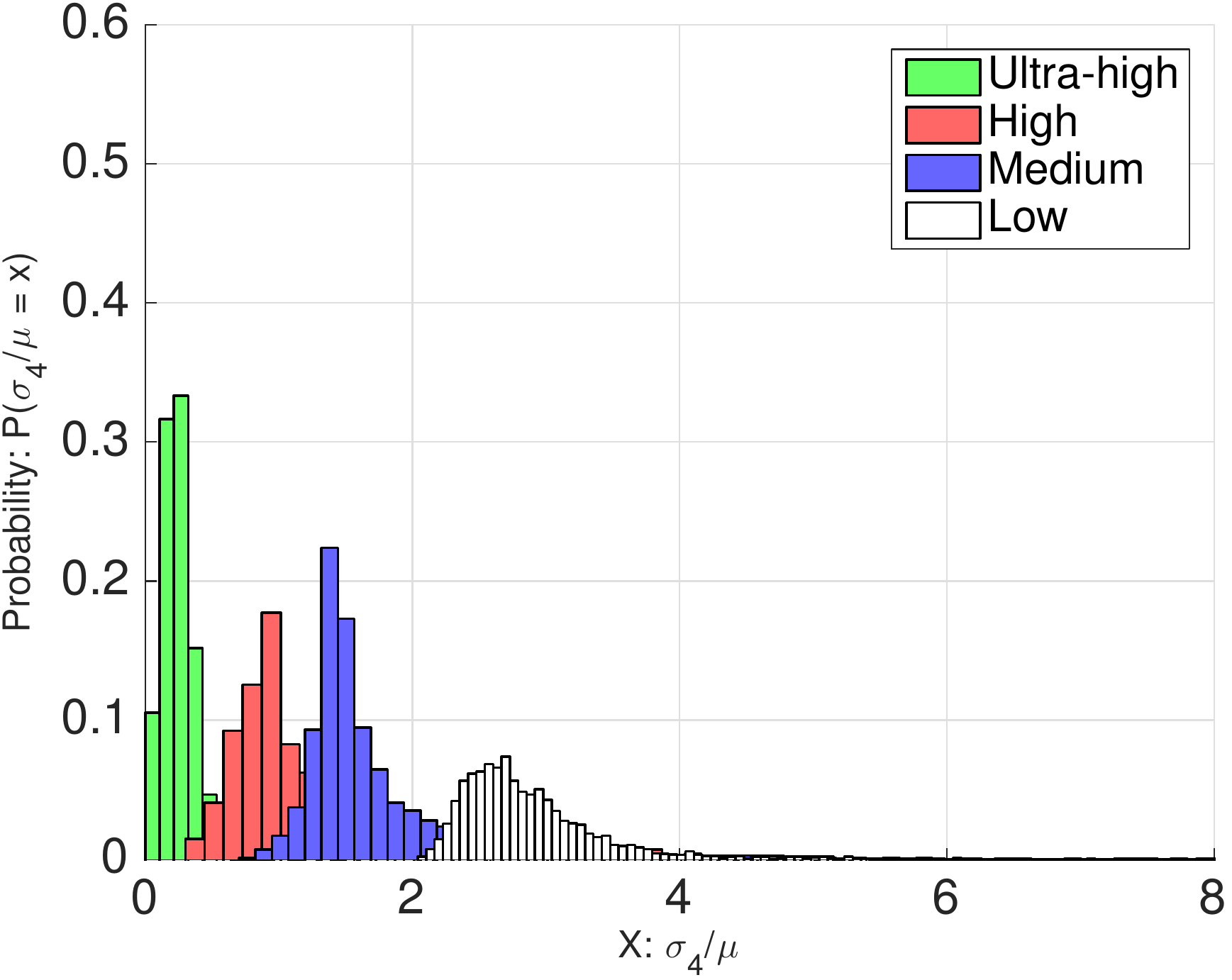}}\quad
				\label{fig:histSig4Vid}
			}
			\hspace{-5mm}
			\subfigure[burstiness at 8-second]{
				{\includegraphics[width=0.32\textwidth,height=.20\textwidth]{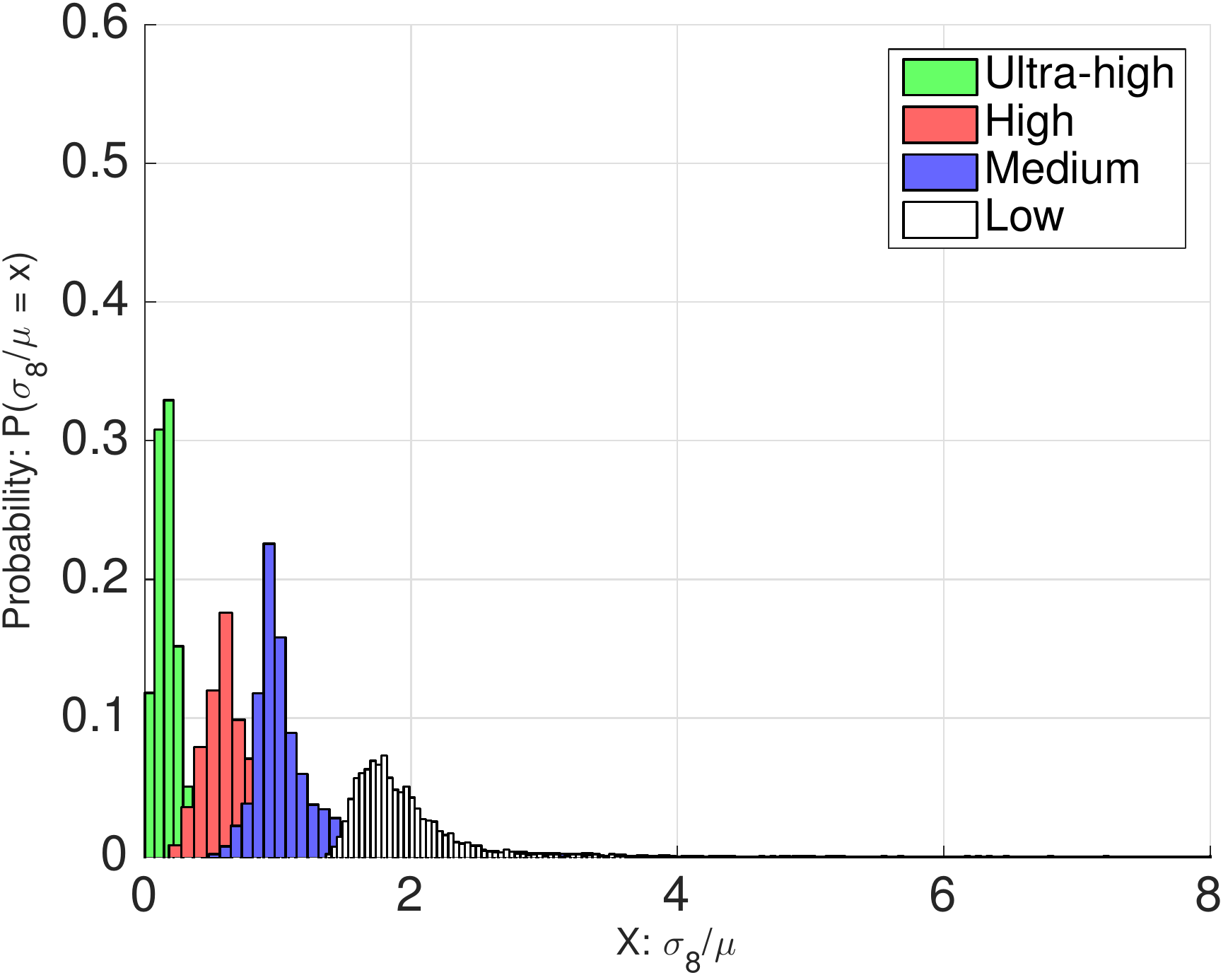}}\quad
				\label{fig:histSig8Vid}
			}
			\hspace{-5mm}
			\subfigure[burstiness at 16-second]{
				{\includegraphics[width=0.32\textwidth,height=.20\textwidth]{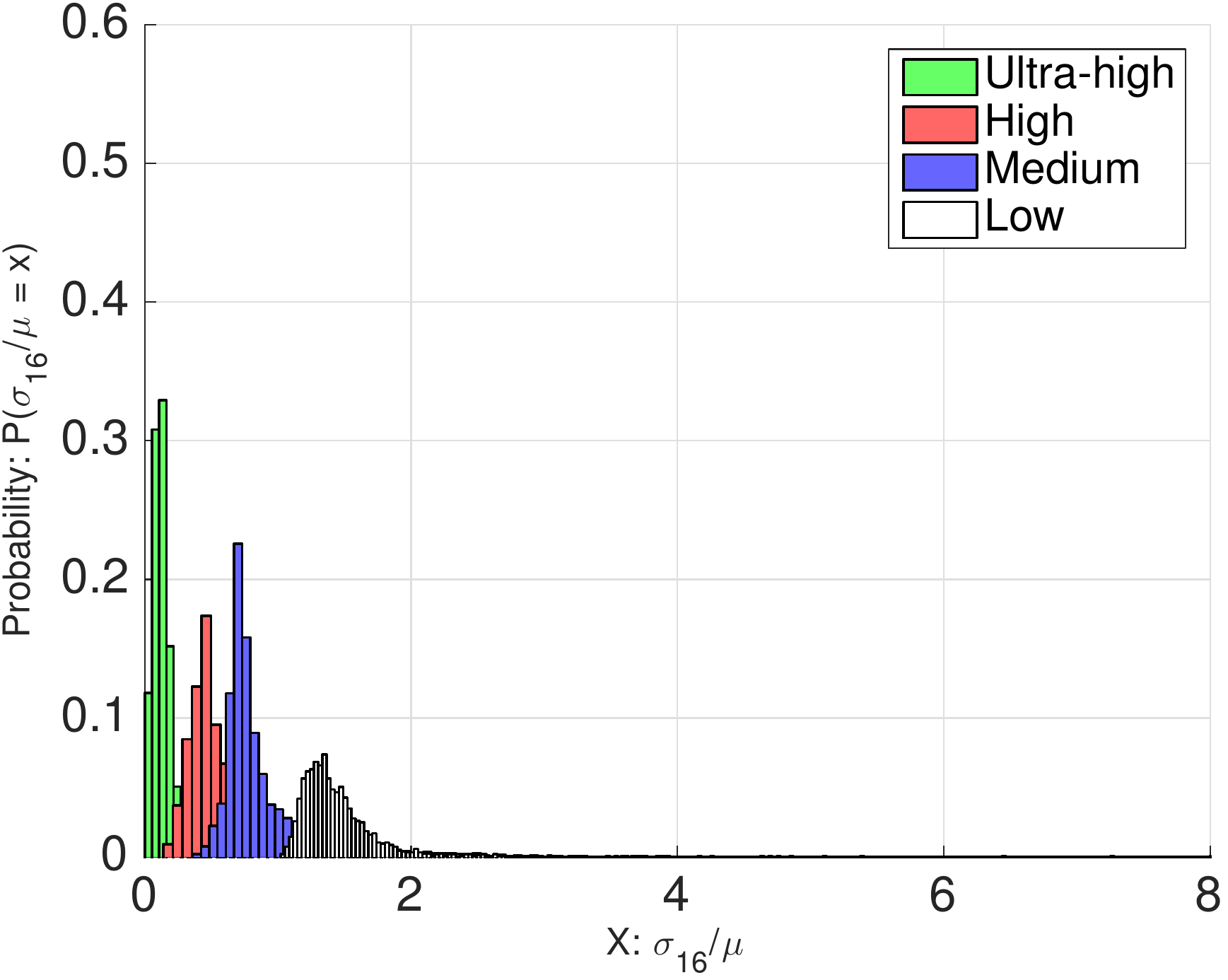}}\quad
				\label{fig:histSig16Vid}
			}
		}				
		\vspace{-4mm}
		\caption{Histogram of idle-time, average rate and burstiness at various time scales (various resolutions). }
		\vspace{-4mm}
		\label{fig:histVid}
	\end{center}
	\vspace{-3mm}
\end{figure*}

\begin{figure}[t!]
	\centering
	\vspace{-2mm}
	\mbox{\includegraphics[width=0.45\textwidth,height=0.20\textheight]{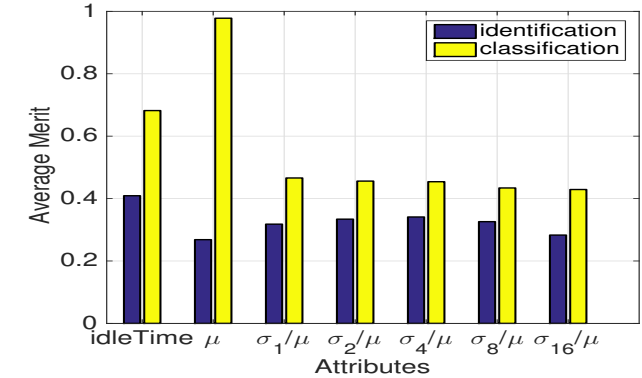}}
	\vspace{-3mm}
	\caption{Merit of attribute.}
	\vspace{-4mm}
	\label{fig:merit}
\end{figure}

The distribution of the attributes used by the resolution classifier is shown in Fig.~\ref{fig:histVid}. The distinctions between the resolutions are again visually apparent: Fig.~\ref{fig:histIdleVid} shows that as video resolution increases, the idle-time fraction distribution (predictably) shifts to the left, whereas the average rate distribution shifts to the right (Fig.~\ref{fig:histRateVid}). The video stream burstiness for the various resolutions is also distinct at the various time-scales, as shown in Figs.~\ref{fig:histSig1Vid}-\ref{fig:histSig16Vid}. 

It is evident from the above that the different attributes will have different importance for the classification engines that identify video streams and their resolutions. In order to quantify the importance, we used the {\tt InfoGain} tool that is part of the Weka Machine Learning package, that determines an average merit score of each attribute based on the measure of information gain from that attribute -- the larger the reduction of entropy (uncertainty), the higher the merit of an attribute. Fig.~\ref{fig:merit} shows the merit of each attribute for the two machines. The blue bars, corresponding to the video identification machine, show that all attributes are nearly equally important, with idle-time being slightly more dominant, and average rate slightly less. By contrast, the video resolution classifier (yellow bars) relies heavily upon the average rate, followed by the idle-time. The burstiness at the various time-scales are roughly equally important to both machines. These merit scores confirm our initial intuition that the selected attributes constitute reasonable inputs to the video classification engines.

\vspace{-3mm}
\subsubsection{Tuning Machine Parameters}
\vspace{-1mm}
We employ three popular classification algorithms: J48, Random Forest, and MLP, from the Weka machine learning library. We tune the parameters of these machine learning algorithms to maximize their performance for the chosen attributes. For example, the Random Forest algorithm has two parameters we can tune -- the depth of the tree and the number of selected attributes for each tree. For each combination of parameters, we evaluated its efficacy via 10-fold cross-validation, whereby the dataset is randomly split into training (90\% of total instances) and validation (10\% of total instances) sets, and accuracy is averaged over 10 runs to produce a single performance metric.

For the video identification machine, the overall accuracy of Random Forest is shown in Table~\ref{tab:parameter} for the various parameter combinations. The highest accuracy (of $95.29$\%) was achieved by setting the tree-depth to $9$ and the number of attributes in the tree to $1$. Increasing tree-depth or number of attributes per-tree beyond these numbers reduces accuracy due to over-fitting. We similarly tuned the J48 algorithm by adjusting the number of instances on each leaf (optimal = $4$), and the MLP algorithm by varying the number of hidden layers (optimal = $8$), in order to maximize performance, yielding accuracy of $95.12$\% and $87.36$\% respectively.

For the video resolution machine, we used a similar method (details omitted due to space constraints) to tune the parameters, and found the optimal settings to be as follows: Random Forest uses a tree-depth of $5$ and number of attributed per-tree of $3$; J48 uses $4$ instances per-leaf; and MLP uses $5$ hidden layers. 

\begin{table}[]
	\caption{Tuning Random-Forest (video identifier.}
	\label{tab:parameter}
	
	\centering
	\scalebox{0.7}{
		\begin{tabular}{|l|l|l|l|l|l|l|l|}
			
			\hline
			\multicolumn{2}{|c}{} &\multicolumn{6}{|c|}{number of attributes}\\ \cline{3-8}
			\multicolumn{2}{|c|}{} & 1       & 2       & 3       & 4       & 5       & 6       \\ \hline
			\multirow{12}{*}{\rotatebox[origin=c]{90}{depth of tree}} &1               & 80.7978 & 89.283  & 91.5748 & 88.077  & 85.7852 & 85.7634 \\ \cline{2-8}
			&2               & 85.5997 & 91.744  & 92.9281 & 93.4083 & 93.7138 & 93.7084 \\ \cline{2-8}
			&3               & 90.7454 & 94.0849 & 94.4996 & 94.576  & 94.5542 & 94.5051 \\ \cline{2-8}
			&4               & 93.7193 & 94.4505 & 94.6033 & 94.7015 & 94.7288 & 94.6306 \\ \cline{2-8}
			&5               & 94.1777 & 94.6797 & 94.8434 & 94.898  & 94.7725 & 94.5869 \\ \cline{2-8}
			&6               & 94.5762  & 94.9089 & 94.9907 & 94.7179 & 94.5978 & 94.6742 \\ \cline{2-8}
			&7               & 95.0071 & 95.0999 & 94.7506 & 94.7397 & 94.7233 & 94.7179 \\ \cline{2-8}
			&8               & 95.1162 & 95.0453 & 94.767  & 94.7124 & 94.6197 & 94.6361  \\ \cline{2-8}
			&9               & \textbf{95.2908} & 94.8981  & 94.767  & 94.7615 & 94.6306 & 94.6579 \\ \cline{2-8}
			&10              & 95.1108 & 94.8052 & 94.707  & 94.7452 & 94.7288 & 94.6906 \\ \cline{2-8}
			&11              & 94.9471 & 94.7831 & 94.6977 & 94.7124 & 94.7233 & 94.6561 \\ \cline{2-8}
			&12              & 94.7834 & 94.7179 & 94.6488 & 94.6397 & 94.7233 & 94.6343 \\ \cline{1-8}
	\end{tabular}}
	\vspace{-2mm}
\end{table}

\begin{figure*}[t!]
	\begin{center}
		\mbox{
			\subfigure[J48]{
				{\includegraphics[width=0.3\textwidth]{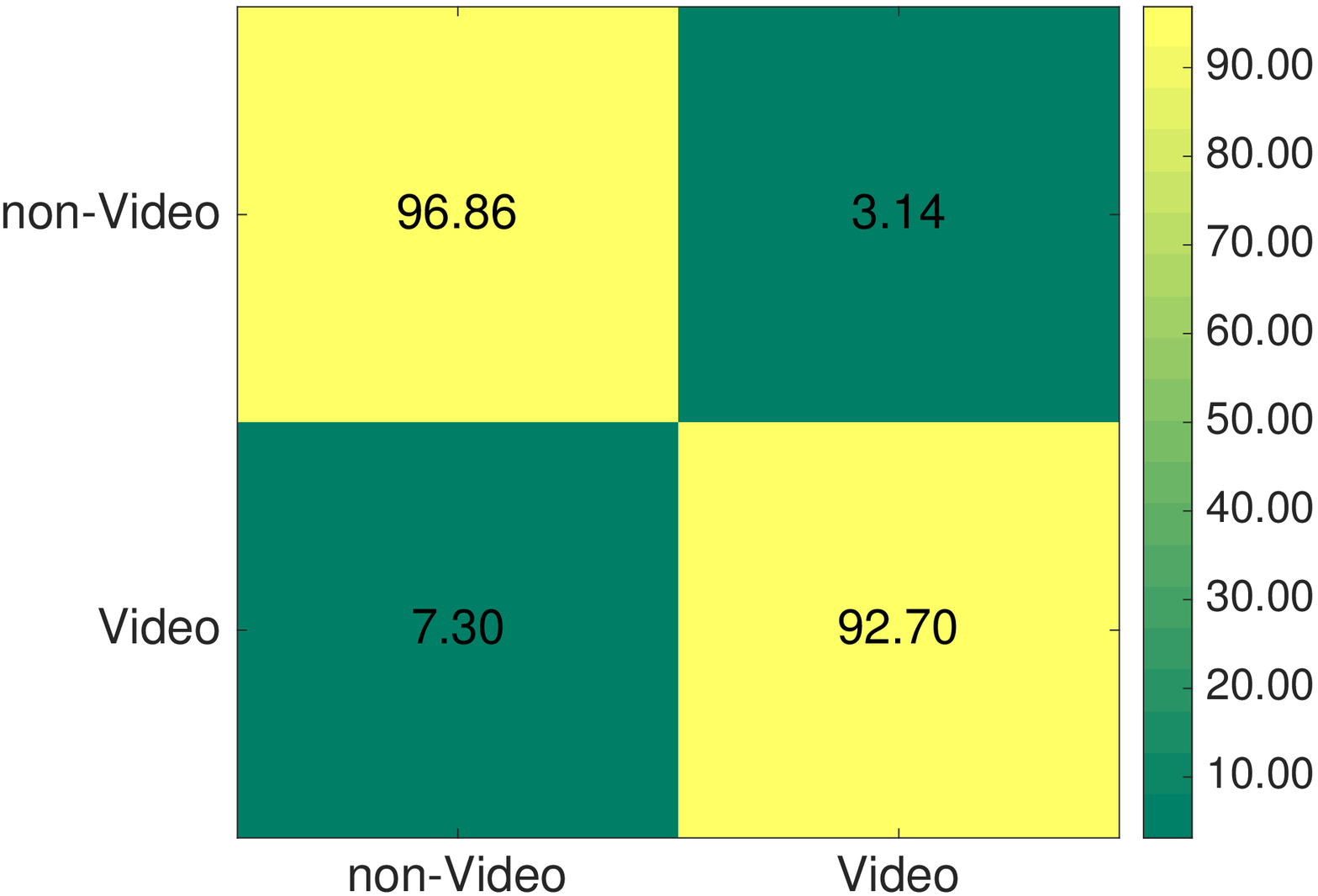}}\quad
				\label{fig:confJ48}
			}
			\hspace{-2mm}
			\subfigure[Random forest]{
				{\includegraphics[width=0.3\textwidth]{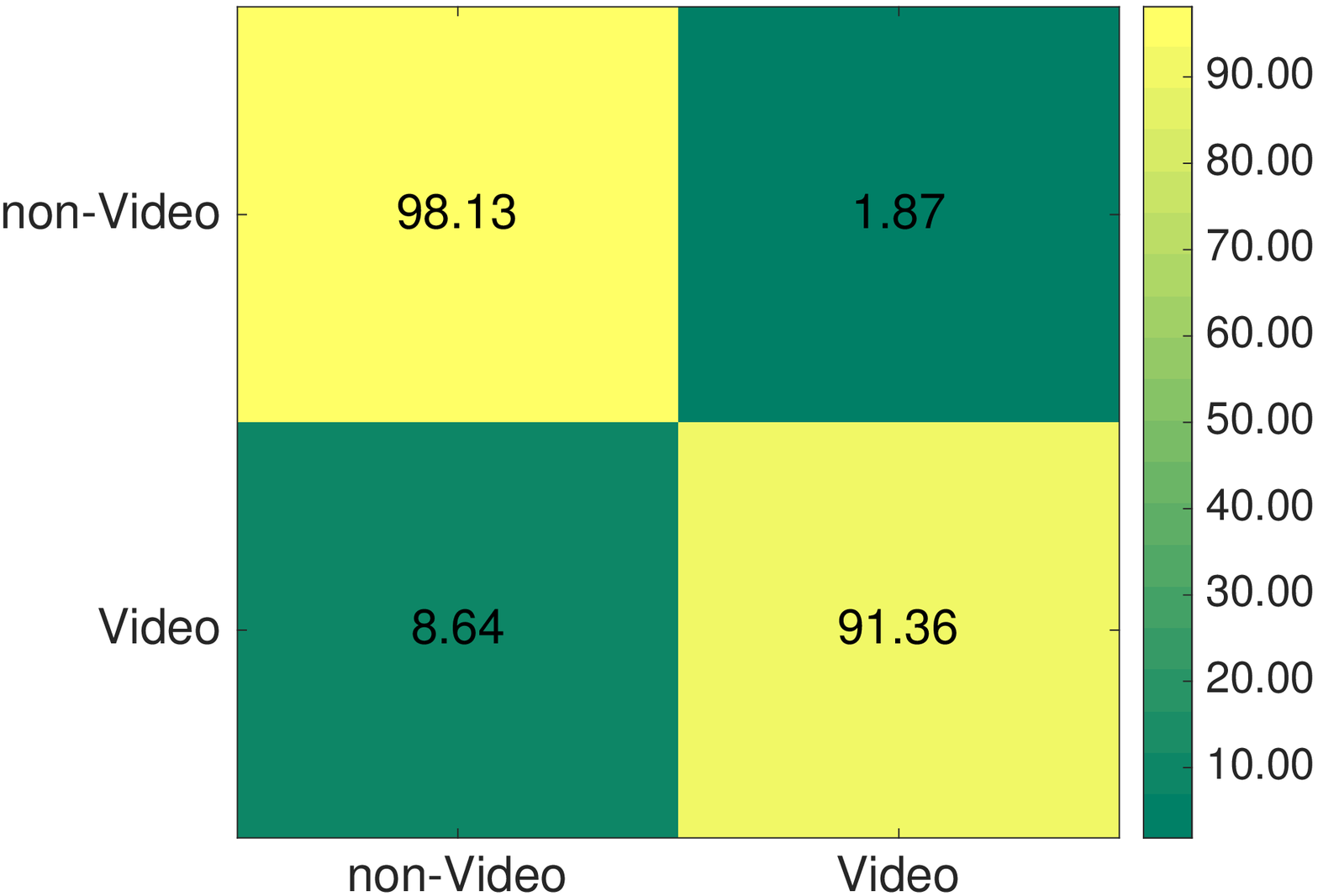}}\quad
				\label{fig:confRandom}
			}
			\hspace{-2mm}
			\subfigure[MLP]{
				{\includegraphics[width=0.3\textwidth]{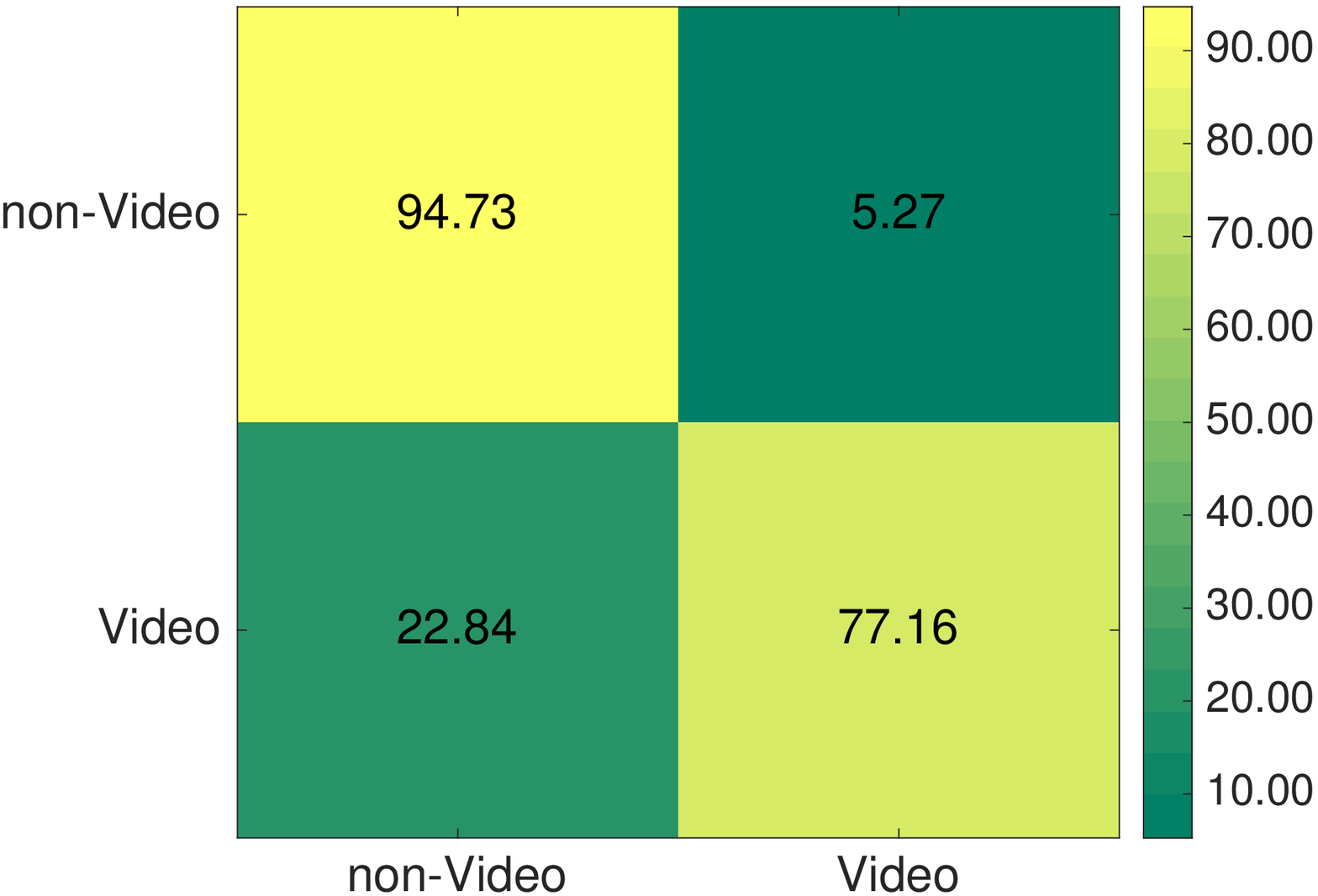}}\quad
				\label{fig:confSVM}
			}
		}
		\vspace{-4mm}
		\caption{Confusion matrix of video identifier}
		\vspace{-5mm}
		\label{fig:confMat}
	\end{center}
\end{figure*}

\begin{figure*}[t!]
	\begin{center}
		\mbox{
			\subfigure[J48]{
				{\includegraphics[width=0.3\textwidth]{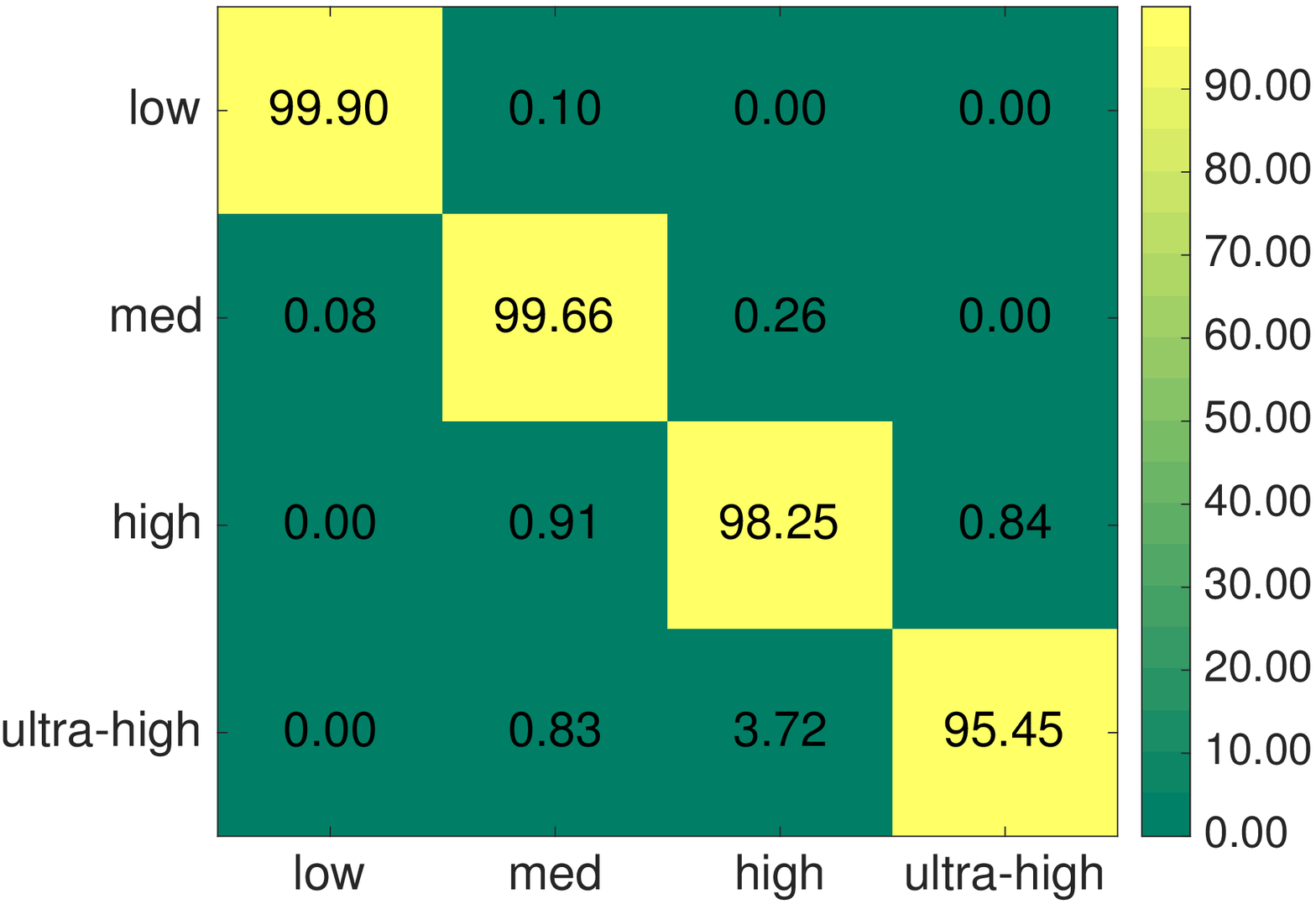}}\quad
				\label{fig:confJ48Vid}
			}
			\hspace{-2mm}
			\subfigure[Random forest]{
				{\includegraphics[width=0.3\textwidth]{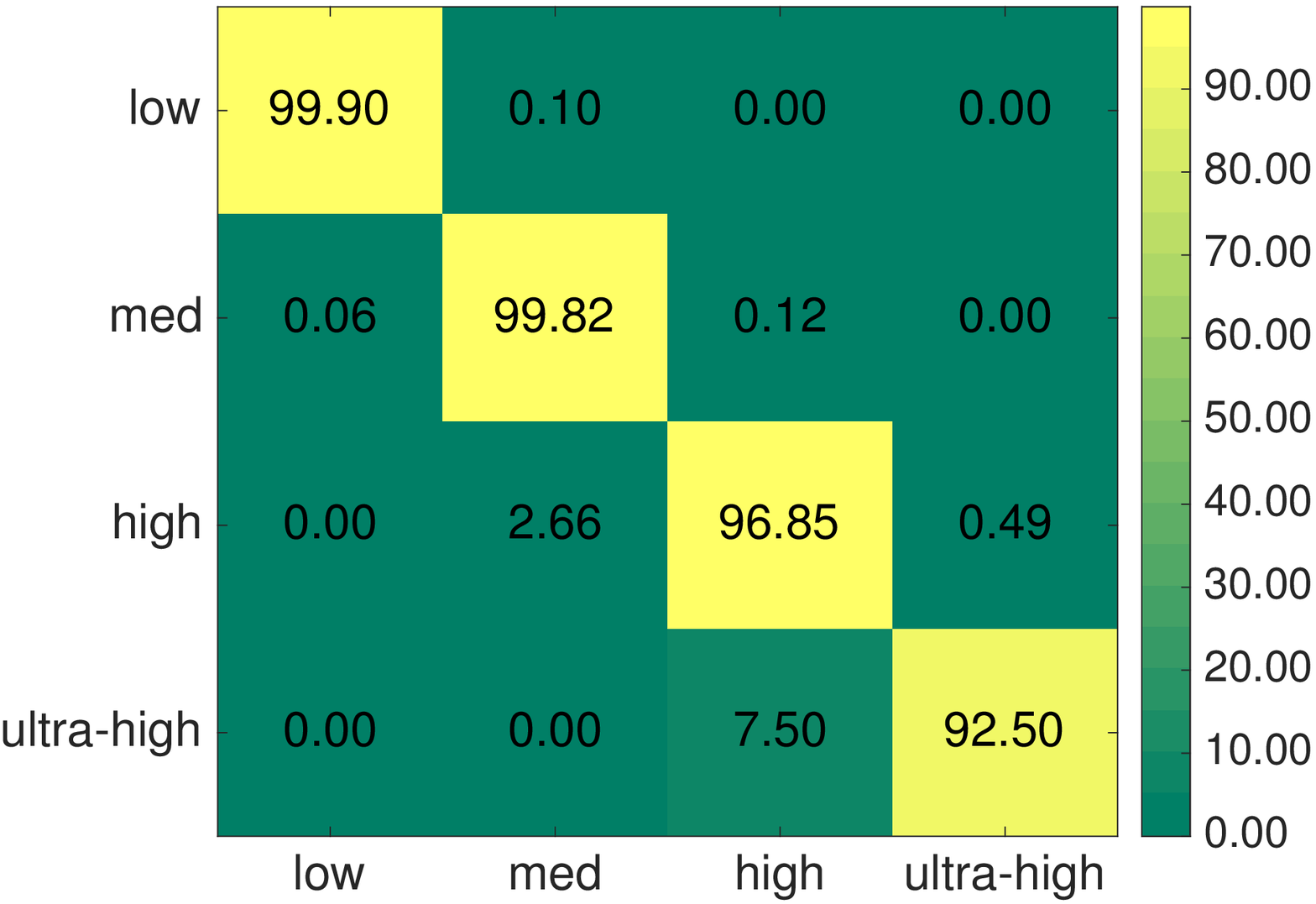}}\quad
				\label{fig:confRandomVid}
			}
			\hspace{-2mm}
			\subfigure[MLP]{
				{\includegraphics[width=0.3\textwidth]{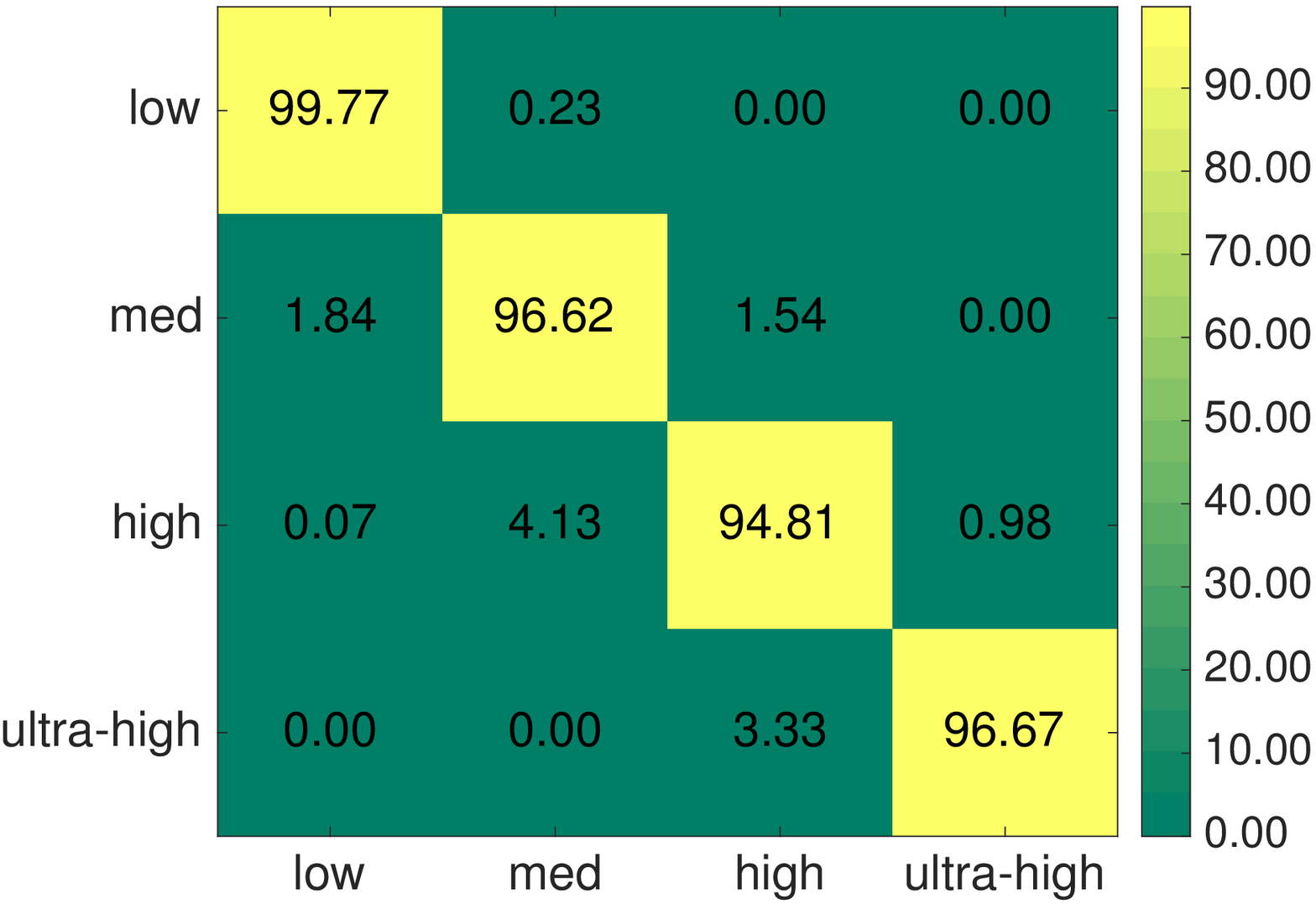}}\quad
				\label{fig:confSVMVid}
			}
		}
		\vspace{-4mm}
		\caption{Confusion matrix of video resolution classifier}
		\vspace{-5mm}
		\label{fig:confMatVid}
	\end{center}
	\vspace{-2mm}
\end{figure*}

\subsubsection{Off-Line Accuracy}
Having tuned the various algorithms to maximize their performance, we now conduct an off-line evaluation of their accuracy on our lab dataset (for which the ground truth is known). A ten-fold cross-validation is performed over the entire dataset, and results are depicted in the form of a confusion matrix, in which rows denote the ground truth and columns the machine output. Fig.~\ref{fig:confMat} shows the accuracy of the video identification machine from the three machine learning algorithms. J48 and Random Forest correctly identify video flows over 90\% of the time, while MLP has a poor true positive rate of 77\%. We believe this is because the geometry of our training instances is more suitable for decision-tree-based classifiers (J48 and Random forest) than for neural-network-based classifiers (MLP). The identification of non-video flows has higher accuracy with all the methods, with J48 and Random Forest being above 96\% and MLP nearly 95\%. We believe the higher false-positive rate for video flows (than non-video flows) is because they can sometimes change their profile (due to network conditions and rebuffering), especially at higher resolutions, making them look closer to downloads. 

Fig.~\ref{fig:confMatVid} shows the confusion matrix for the resolution classifier. All three machine learning models yield a fairly high accuracy, being over $99$\% at low resolutions, and dropping somewhat at higher resolutions. J48 and Random Forest outperform MLP at higher resolutions, though they all tend to sometimes classify ultra-high resolution videos as high resolution, and medium resolution videos as high resolution. These errors (of about a few percent) are not surprising, since the attributes of the high resolution videos overlap with those of medium and ultra-high resolution flows, as depicted in Fig.~\ref{fig:histVid} -- specifically, the two most important attributes, idle-time and average rate, have significant overlaps.

\begin{figure*}[t!]
	\begin{center}
		\mbox{
			\subfigure[Video identification]{
				{\includegraphics[width=0.43\textwidth,height=.25\textwidth]{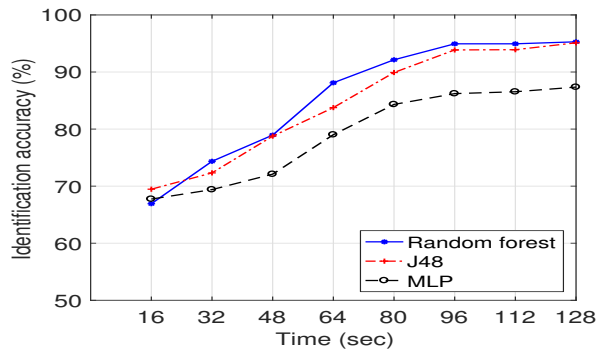}}\quad
				\label{fig:id}
			}
			\subfigure[Resolution classification]{
				{\includegraphics[width=0.43\textwidth,height=.25\textwidth]{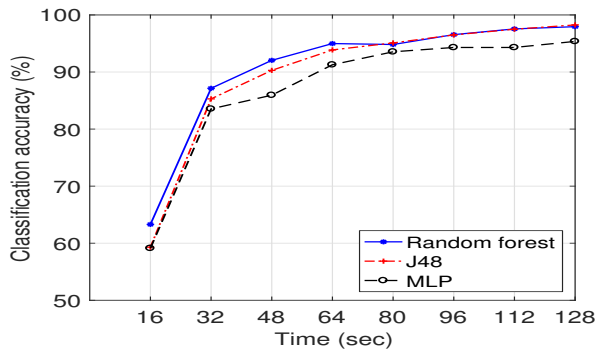}}\quad
				\label{fig:cls}
			}
		}
		\vspace{-4mm}
		\caption{Real-time accuracy of (a) video identification, and (b) video resolution classification. }
		\vspace{-2mm}
		\label{fig:accuracy}
	\end{center}
	\vspace{-3mm}
\end{figure*}

\vspace{-3mm}
\subsubsection{Real-Time Accuracy}
Recall that certain attributes (such as burstiness at time-scales of 8 and 16 seconds) become available only after the flow has been active for a certain duration. We now evaluate the accuracy of our classification methods when they operate in real-time, namely as and when flow ``sub-profiles'' become available from the first 16 seconds to the past one minute over their two-minute lifespan. Fig.~\ref{fig:id} shows the time evolution of real-time classification accuracy -- video streams are identified with an accuracy of about 60\% if only the first 16 seconds worth of their profile is available to the classifier. This is also because video flows tend to buffer in the beginning, which makes them less distinguishable from downloads in the initial few seconds. As the length of sub-profiles increases, so does the accuracy -- after 48 seconds, 80\% accuracy is achieved, and this rises to 95\% for J48 and Random Forest at about a minute-and-a-half. 

Similarly, Fig.~\ref{fig:cls} shows that the accuracy of the resolution classifier increases rapidly with the length of sub-profile. This is not surprising, as various attributes computed during the first 16 seconds do not perfectly identify/classify video flows due to their initial buffering. For example, an ultra-high resolution video (Fig.~\ref{fig:prof2160}) is very similar to a large download if one considers the idle-time, average rate and burstiness for only the initial 16 or 32 seconds of the profile. The attributes $\sigma_{8}/\mu$ and $\sigma_{16}/\mu$ become available respectively only after 32 and 64 seconds of stream activity, and are fairly important for the classification, leading to a very rapid rise in accuracy at around the minute mark.

\textbf{Summary}: Our system uses a judicious combination of software and hardware to isolate elephant flows and monitor their individual behavior. Flow attributes such as idle-time fraction, average rate, and burstiness at various time-scales are extracted without packet-level inspection, and fed to a machine learning model. Video flows can be identified by our machine with 70\% certainty within 30 seconds, rising to over 90\% within two minutes, while video resolution can be correctly deduced with 80\% accuracy in 30 seconds, rising to over 95\% within two minutes. In what follows we evaluate the scalability of our system and discuss insights obtained from a real deployment.

\vspace{-4mm}
\section{Evaluation and Deployment}\label{eval}

We briefly evaluate the scalability of our system to large flow numbers and high arrival rates (\S\ref{eval:perf}), and describe insights obtained from a trial deployment (\S\ref{eval:dorm}).

\subsection{Scalability Test}\label{eval:perf}
We subject our system to stress-testing with emulated elephant flows from a traffic generator. We connect a Spirent \cite{Spirent} TestCenter chassis SPT-11U (firmware v4.24.1026), which is a high-precision commercial-grade hardware traffic generator equipped with a 12-port GE HyperMetric test module, to our NoviSwitch on two ports. One port of the Spirent generates traffic streams representative of video servers, while the other port receives traffic back from our system to represent end-user clients. We wrote TCL scripts to automate the process of traffic emulation: 14 pairs of transmitter/receiver were created, each allocated a distinct /28 public-IP address, and each pair establishes 20 parallel stream blocks each of a separate layer-4 (TCP) port number, thereby generating 280 concurrent flows per second. Further, the port number of each stream block kept circulating each second over a range of 114, resulting in a total of 31920 flows. Each flow sends traffic at a constant rate uniformly distributed between [0.8, 1.2] Mbps (representative of a 360p video). The emulation was run for 300 sec.

\begin{figure}[!t]
	\centering
	\vspace{-5mm}
	\includegraphics[width=0.42\textwidth,height=0.17\textheight]{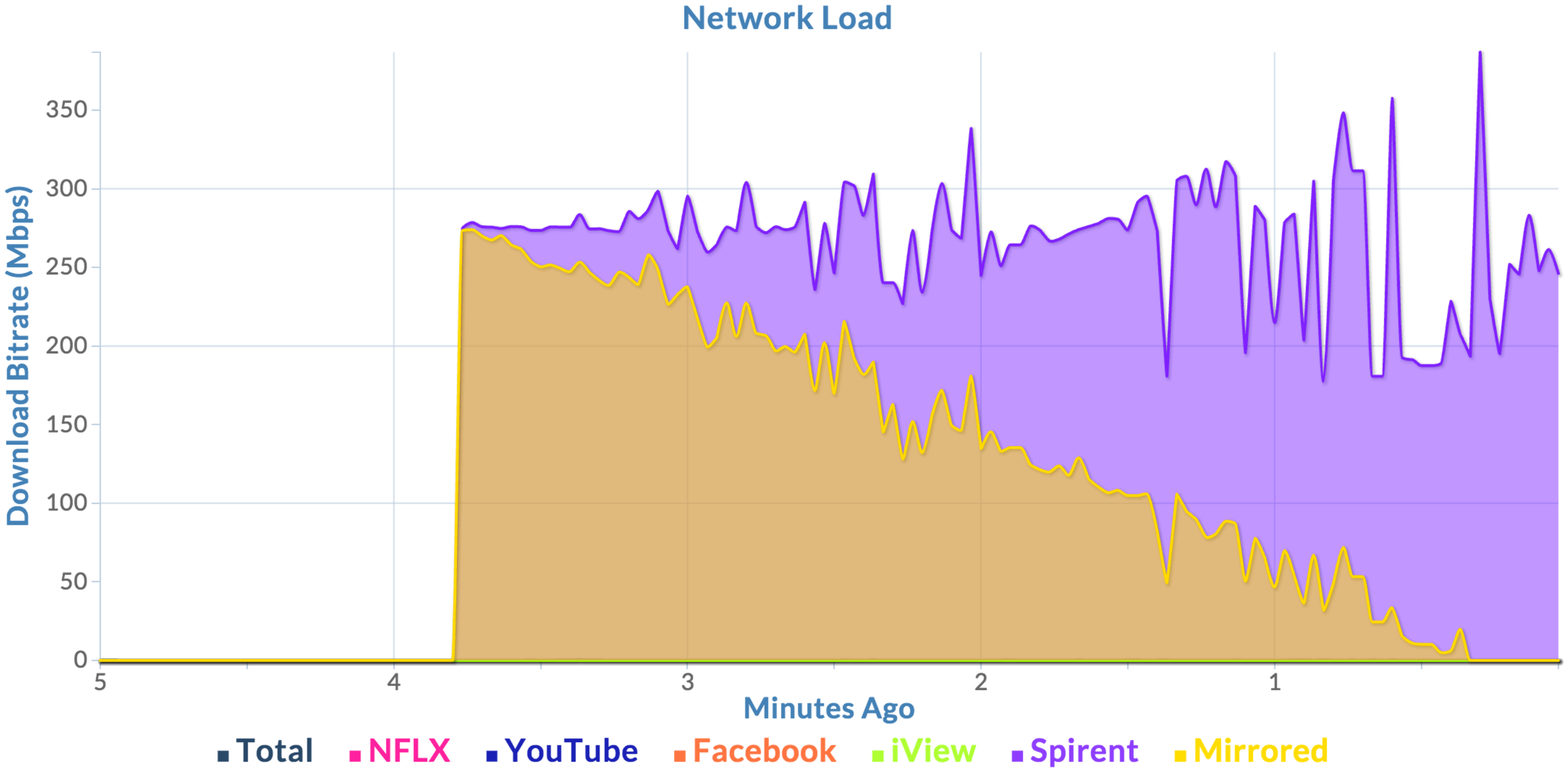}
	\vspace{-8mm}
	\caption{Network load (31920 flows arrive at the rate of 280 flows-per-second).}
	\label{fig:sp_load_normal}
\end{figure}

\begin{figure}[!t]
	\centering
	\vspace{-15mm}
	\includegraphics[width=0.42\textwidth]{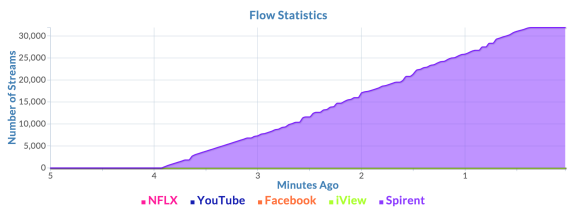}
	\vspace{-15mm}
	\caption{Flow statistics (31920 flows arrive at the rate of 280 flows-per-second).}
	\label{fig:sp_flow_normal}
\end{figure}

Fig.~\ref{fig:sp_load_normal}  depicts the link load (purple line) and mirror load (brown line) at 1s intervals. The measured loads (deduced from OpenFlow counters) corroborate very well with the actual loads (reported by Spirent), being within 1.7\% of each other, confirming that our system measures flow rates accurately. Further, the mirror traffic load (sent for software processing) is initially at 100\% of offered load, but gradually drops to zero (over a period of 210 seconds) as the reactive flow entries are pushed into the OpenFlow switch to stop the mirroring of elephant flows. Fig.~\ref{fig:sp_flow_normal} shows the ramp-up in the number of reactive flow-table entries, being pushed at the rate of around 280 flow-mods per-second. The stress-test is meant to ensure that our system is scalable to large number of active elephant flows (31920 in this case), and to handle high rate of new flows in the switch hardware (280 new flows per-second), ensuring proper operation in real networks, as described next. 

\subsection{Campus Deployment}\label{eval:dorm}
Our iTeleScope system has been deployed and operational for several months in a university network serving hundreds of students resident in the on-campus dorm. The University IT department provisioned a full feed of the dorm traffic to our system, and we obtained ethics clearance (approval number will be disclosed when this paper is de-anonymized) from our organization in order to conduct this trial, since it gives us access to all network traffic. 

\begin{figure*}[t!]
	\begin{center}
		\mbox{
			\subfigure[Total num. of video streams]{
				{\includegraphics[width=0.20\textwidth]{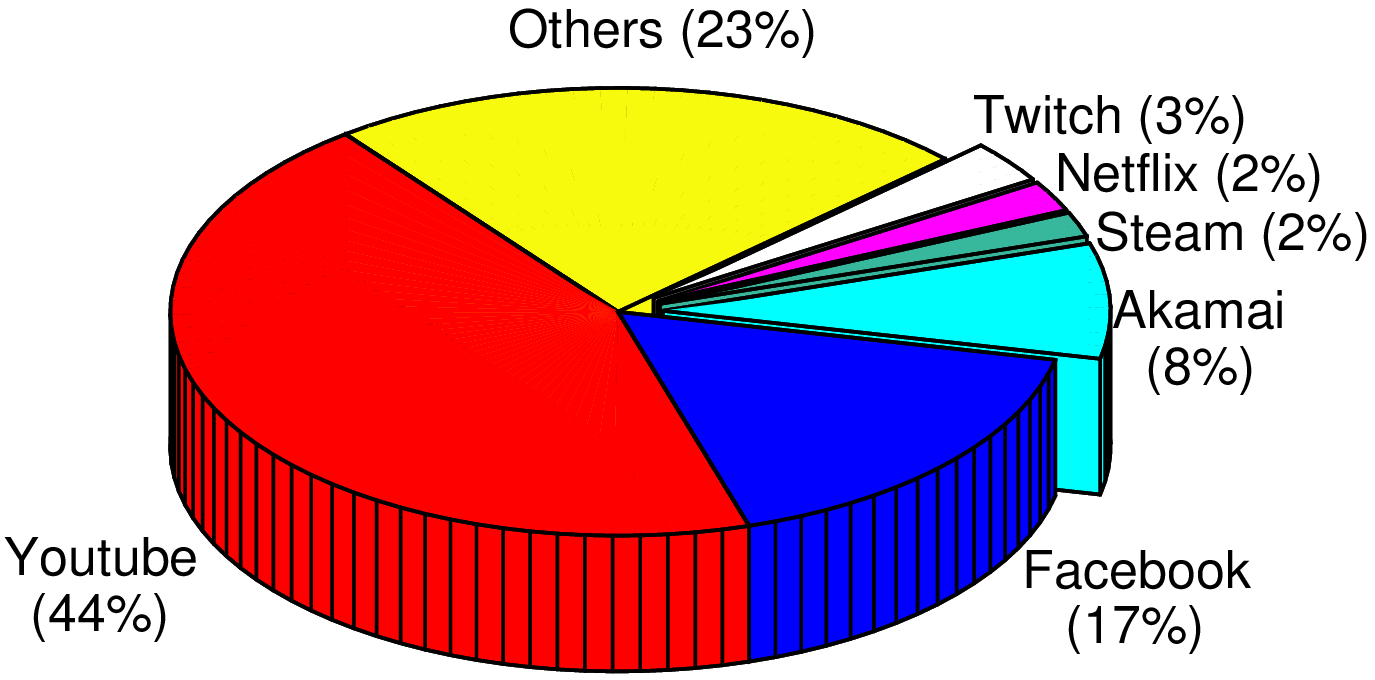}}\quad
				\label{fig:dormPie}
			}
			\hspace{-4mm}
			\subfigure[Daily num. of video streams]{
				{\includegraphics[width=0.30\textwidth]{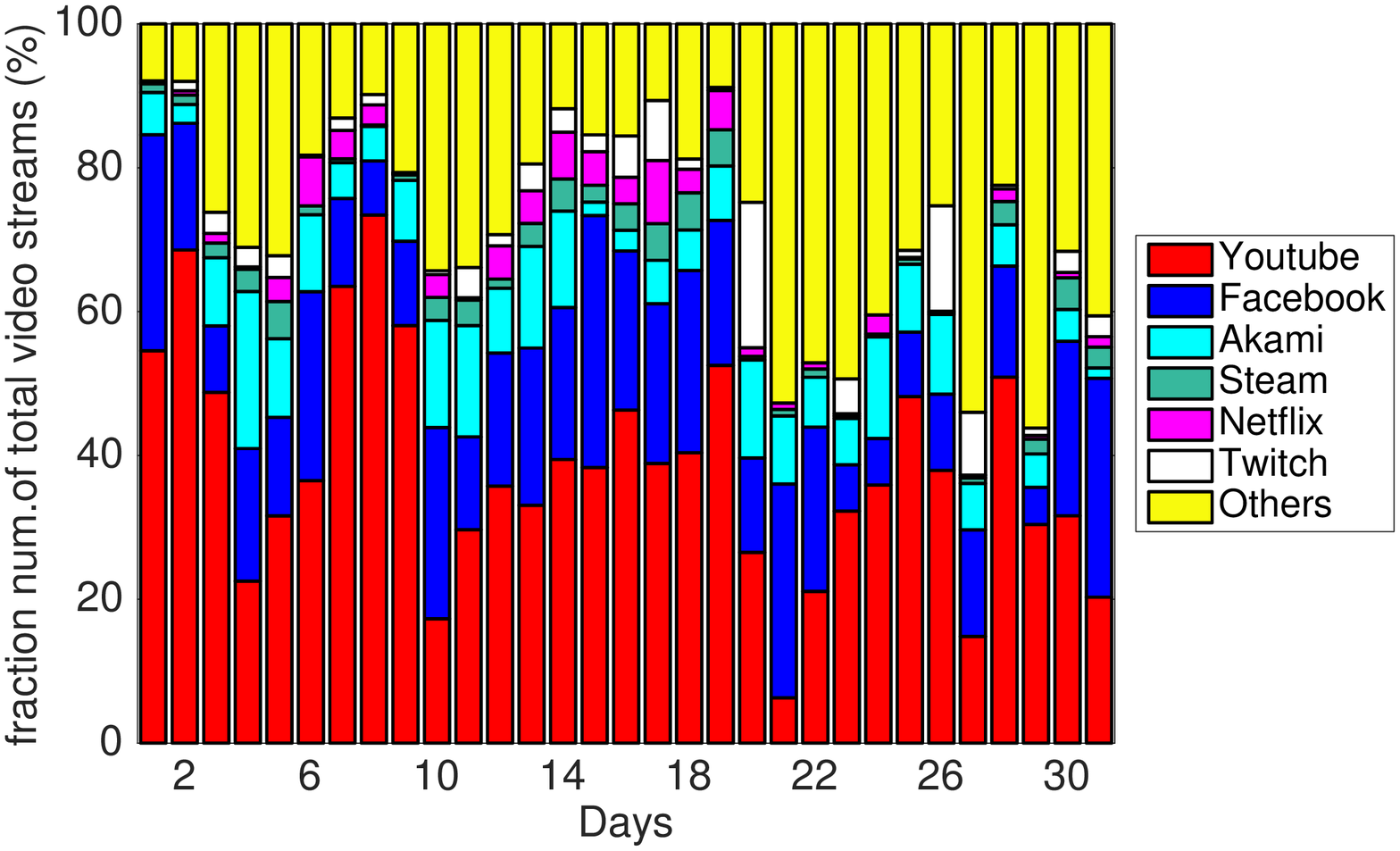}}\quad
				\label{fig:dormDaily}
			}
			\hspace{-6mm}
			\subfigure[Hourly num. of video resolution]{
				{\includegraphics[width=0.23\textwidth]{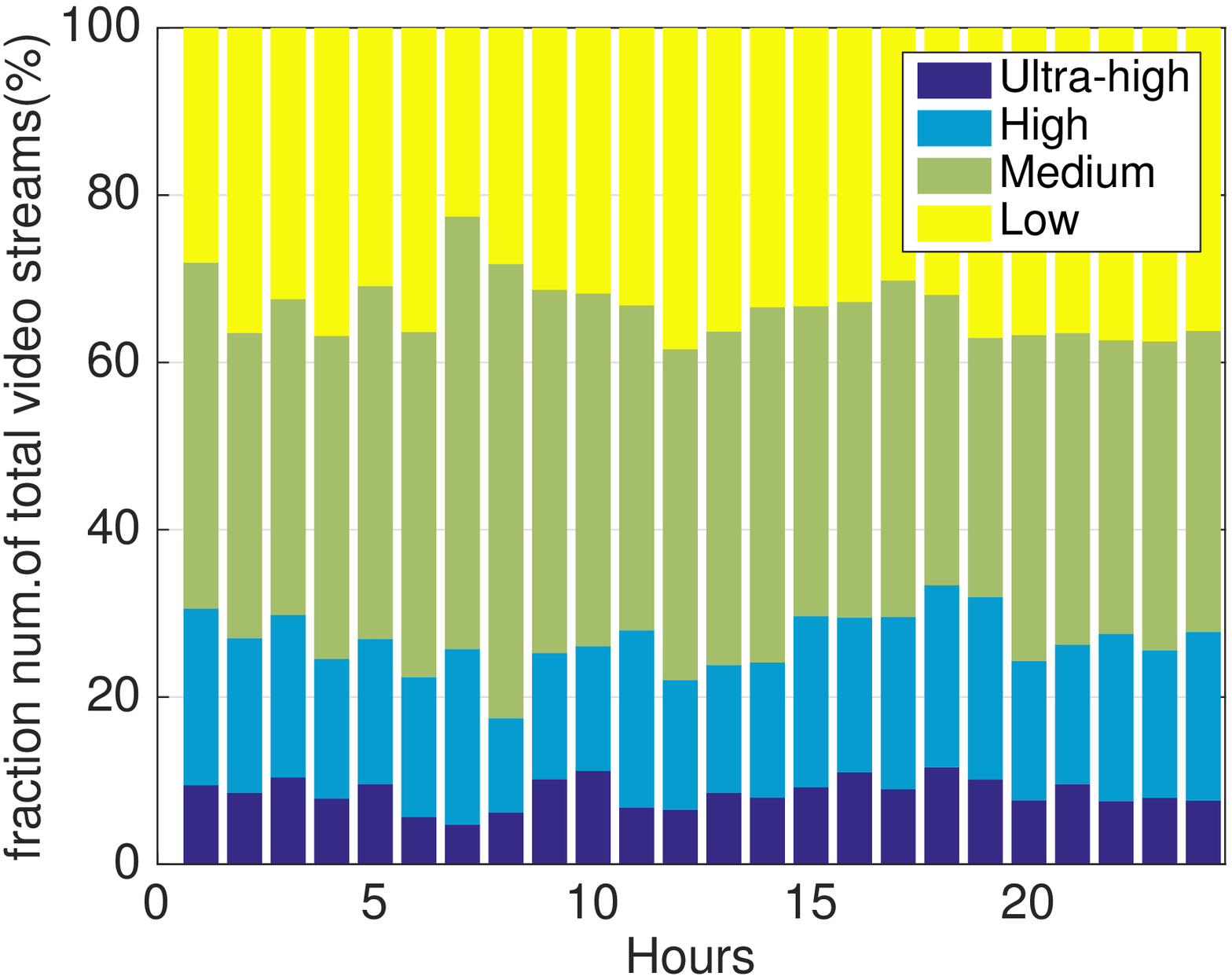}}\quad
				\label{fig:dormRES}
			}
			\hspace{-4mm}
			\hfill
			\subfigure[CCDF of resolution change]{
				{\includegraphics[width=0.225\textwidth]{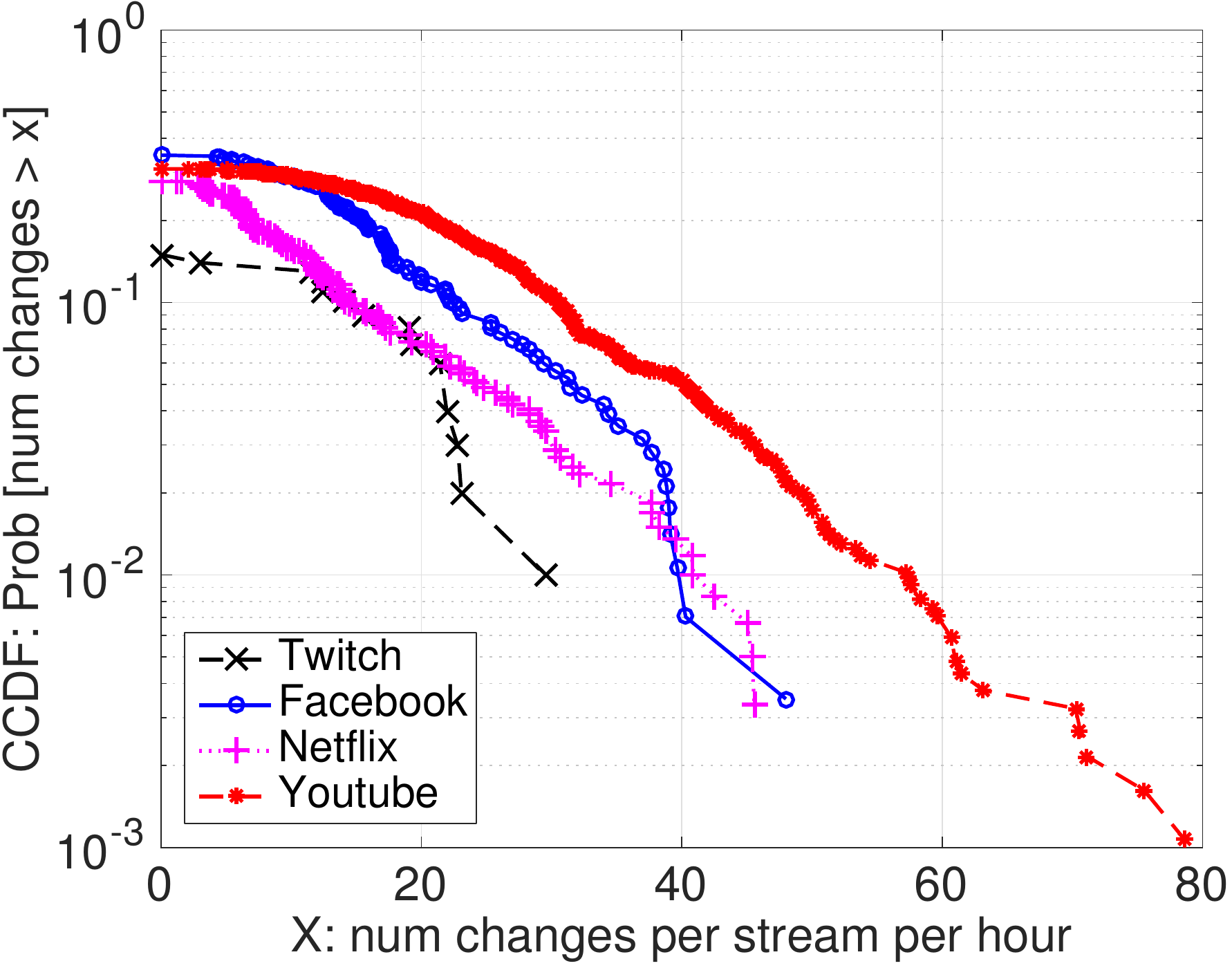}}\quad
				\label{fig:dormResChange}
			}
		}
		\vspace{-4mm}
		\caption{Distribution of Dorm Video Consumption.}
		\vspace{-3mm}
		\label{fig:dorm}
	\end{center}
	\vspace{-2mm}
\end{figure*}

Our system not only displays video flow information (end-point, provider, duration, rate, resolution) in real-time (the live URL will be released when this paper is de-anonymized), but also records video flow information into an InfluxDB that can be analysed post-facto. In what follows we highlight some of the insights we obtained from our system's flow database over a one month period during the academic term. In Fig.~\ref{fig:dormPie} we show a pie-chart of the fraction of streams from the most popular video content providers -- not surprisingly, Youtube and Facebook video streams dominante at 44\% and 17\% respectively. The gaming video platform Twitch contributes 3\% of streams, more than Netflix (2\%), most likely because students tend to prefer free over paid content. Around 8\% of video flows are sourced from Akamai media servers (i.e. \verb|akamai.net| and \verb|akamaiedge.net|). Lastly, our system allowed identification of many other video providers such as Tencent, Youku, Amazon, Yahoo, Instagram, Fastly, Alibaba, Baidu, Huya, Battlenet, HLtv, OurDvs, and Dailymotion (grouped under ``Others'' in Fig.~\ref{fig:dormPie}) that collectively constitute 23\% of video streams during the month. This break-down of video streams by provider elicited much interest from the IT department, who had no prior visibility into video viewing patterns (especially for less popular video providers) in the campus dorm.

\begin{figure}
	\subfigure[Video duration]{
		{\includegraphics[width=0.49\textwidth]{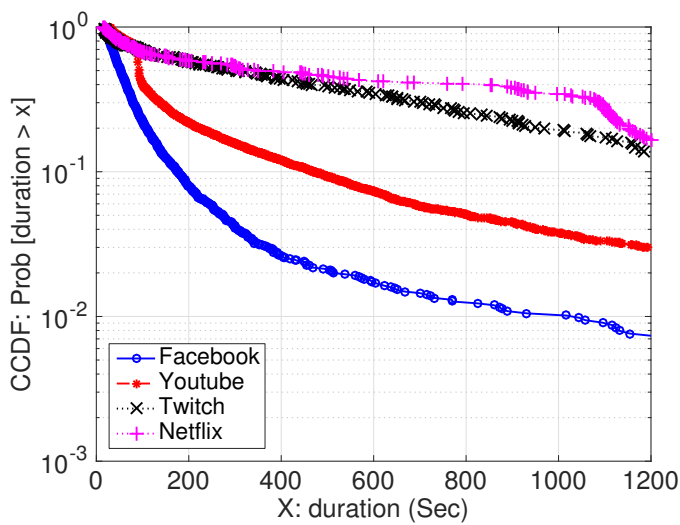}}
		\label{fig:dormVideoDuration}
	}%
	\hfill
	\hspace{-2mm}
	\subfigure[Video rate]{
		{\includegraphics[width=0.49\textwidth]{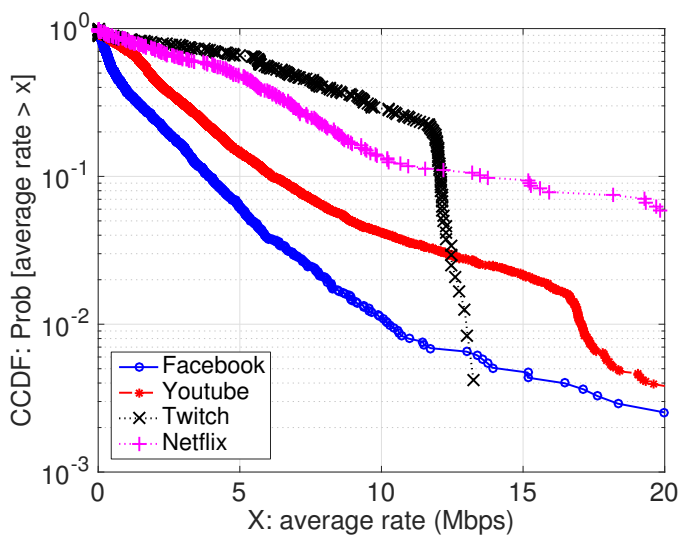}}
		\label{fig:dormVideoRate}
	}
	\caption{CCDF of Dorm Video characteristics.}
	\label{fig:dormCCDF}
\end{figure}

The day-by-day video consumption pattern over the month is shown in Fig.~\ref{fig:dormDaily}. It is seen that there is substantial fluctuation in the relative proportion of video providers from day to day, and interestingly, the dorm residents tend to watch Twitch gaming videos more on weekends than weekdays. In Fig.~\ref{fig:dormRES} we plot the fraction of video streams at different resolutions on an hourly basis (averaged over the selected month). Surprisingly, a majority of videos are playing at medium resolution and only a small fraction of videos are at ultra-high resolution, though the campus network has abundant bandwidth and rarely experiences congestion. We believe that this is because most of the free content on Youtube and Facebook is only available at medium or lower resolution (i.e. 144p, 240p, 360p, 480p and 720p).

Our system also lets us analyze mid-stream resolution changes -- in Fig.~\ref{fig:dormResChange} we plot the CCDF of resolution changes (normalized to a per-hour basis) in video streams from Youtube, Facebook, Netflix and Twitch. Unsurprisingly, Youtube videos are the most aggressive in adapting their resolution, with $20$\% of streams adapting their resolution at least 20 times per hour (i.e. once every 3 minutes on average), and $1$\% of streams adapting once per minute. Facebook videos are generally shorter (more on this next), yet $11$\% adapt their resolution on average every 3 minutes. Netflix videos tend to adapt their resolution less frequently, with only $7$\% of videos changing their resolution every 3 minutes on average. Twitch videos show the least adaptation, with $85$\% of them never changing their resolution during their entire playback (which can be reasonably long).

Further insights into the dorm video viewing patterns are shown in Fig.~\ref{fig:dormCCDF}, depicting the CCDFs of playback duration and average bit-rates for the 4 popular content providers (Facebook, Youtube, Twitch, and Netflix) over the selected month. Fig.~\ref{fig:dormVideoDuration} confirms that Netflix and Twitch videos are watched for reasonably long durations (nearly 40\% of streams last longer than 10 minutes), followed by Youtube and Facebook for which 7\% and 2\% of videos are respectively watched for longer than 10 minutes by dorm residents. The average bit-rates shown in Fig.~\ref{fig:dormVideoRate} also confirm that Twitch and Netflix videos are more bandwidth intensive than Youtube and Facebook videos -- Twitch and Netflix use on average 6.6 Mbps while this measure is 2.8 Mbps and 1.5 Mbps for Youtube and Facebook respectively. 

Our system is being trialled in a Tier-1 carrier network, and is providing real-time video traffic visibility as well as off-line reporting of video consumption patterns; the outputs are comparable to those from a commercial DPI appliance, but at a fraction of the cost. Confidentiality requirements unfortunately prevent us from disclosing any findings from the trial.

\section{Conclusion}\label{con}
Video traffic dominates enterprise and carrier network traffic, yet operators have limited visibility into the number, duration, and quality of video flows traversing their network. Existing solutions are either hardware-based and expensive, or software-based and unscalable. Our solution, iTelescope, judiciously combines software packet-level inspection with hardware flow-level telemetry to isolate elephant flows and extract flow attributes, that are used in conjunction with machine learning to identify and classify video flows in real-time at low-cost. We have built our solution using off-the-shelf SDN hardware and open-source software. We have trained and validated the accuracy of our machine learning algorithms in the lab, demonstrated our system scalability to tens of thousands of concurrent streams, and deployed it in a live network serving hundreds of real users. Our solution provides unprecedented visibility to network operators, and has the potential to become a platform for actively managing video delivery quality on a per-stream basis in the near future.

\balance
{\footnotesize \bibliographystyle{acm}
\bibliography{iTeleScope}}

\end{document}